\documentclass[onecolumn,amsmath,amssymb,11pt,nofootinbib]{revtex4}
\pdfoutput=1
\usepackage[bookmarks,linktocpage, colorlinks=true, plainpages = false, citecolor = blue,  linkcolor=blue, urlcolor = blue,
filecolor = blue]{hyperref}

\usepackage{graphicx}
\usepackage{dcolumn}
\usepackage{bm}
\usepackage{braket}
\usepackage{verbatim} 
\usepackage[]{inputenc}
\usepackage{cancel}
\usepackage[usenames,dvipsnames]{color}
\usepackage{natbib}

\def\gsim{ \lower .75ex \hbox{$\sim$} \llap{\raise .27ex \hbox{$>$}} }
\def\lsim{ \lower .75ex \hbox{$\sim$} \llap{\raise .27ex \hbox{$<$}} }

\newcommand{\nn}{\nonumber}
\newcommand{\be}{\begin{equation}}
\newcommand{\ee}{\end{equation}}
\newcommand{\ba}{\begin{aligned}}
\newcommand{\ea}{\end{aligned}}
\newcommand{\bea}{\begin{eqnarray}}
\newcommand{\eea}{\end{eqnarray}}
\renewcommand{\d}{\mathrm{d}}

\def\de{\delta}

\def\c{\chi}

\def\s{\sigma}

\def\ep{\epsilon}
\def\f{\phi}

\def\m{\mu}

\def\t{\tau}

\def\p{\partial}
\def\mR{\mathcal{R}}
\def\mN{\mathcal{N}}
\def\tf{\tfrac}
\def\fr{\frac}

\begin{document}

\allowdisplaybreaks

\begin{titlepage}

\title{Converting entropy to curvature perturbations after a cosmic bounce \vspace{.3in}}

\author{Angelika Fertig}
\email[]{angelika.fertig@aei.mpg.de}
\affiliation{Max Planck Institute for Gravitational Physics \\ (Albert Einstein Institute), 14476 Potsdam-Golm, Germany}

\author{Jean-Luc Lehners}
\email[]{jean-luc.lehners@aei.mpg.de}
\affiliation{Max Planck Institute for Gravitational Physics \\ (Albert Einstein Institute), 14476 Potsdam-Golm, Germany}

\author{Enno Mallwitz}
\email[]{enno.mallwitz@aei.mpg.de}
\affiliation{Max Planck Institute for Gravitational Physics \\ (Albert Einstein Institute), 14476 Potsdam-Golm, Germany}

\author{Edward Wilson-Ewing}
\email[]{wilson-ewing@aei.mpg.de}
\affiliation{Max Planck Institute for Gravitational Physics \\ (Albert Einstein Institute), 14476 Potsdam-Golm, Germany}

\begin{abstract}

\vspace{.3in} \noindent
We study two-field bouncing cosmologies in which primordial perturbations are created in either an ekpyrotic or a matter-dominated contraction phase. We use a non-singular ghost condensate bounce model to follow the perturbations through the bounce into the expanding phase of the universe. In contrast to the adiabatic perturbations, which on large scales are conserved across the bounce, entropy perturbations can grow significantly during the bounce phase. If they are converted into adiabatic/curvature perturbations {\it after} the bounce, they typically form the dominant contribution to the observed temperature fluctuations in the microwave background, which can have several beneficial implications. For ekpyrotic models, this mechanism loosens the constraints on the amplitude of the ekpyrotic potential while naturally suppressing the intrinsic amount of non-Gaussianity. For matter bounce models, the mechanism amplifies the scalar perturbations compared to the associated primordial gravitational waves. 

\end{abstract}

\maketitle

\end{titlepage}

\tableofcontents

\section{Introduction}

The currently accepted paradigm for the formation of all structure in the universe is that stars and galaxies collapsed from small primordial density perturbations. Direct evidence for the existence of such early density perturbations is provided by the temperature fluctuations in the cosmic microwave background. But what caused these primordial density perturbations? An intriguing possibility (and currently the theoretically best studied option) is that these density fluctuations arose from the amplification of quantum perturbations via particular cosmological dynamics. The most popular version of this scenario involves a period of accelerated expansion, inflation, preceding the hot big bang cosmological evolution \cite{Mukhanov:1981xt, Hawking:1982cz, Guth:1982ec, Starobinsky:1982ee, Bardeen:1983qw}. But this is not the only possibility. Quantum perturbations can also be amplified during phases of cosmological contraction, in particular during an ekpyrotic phase of slow high-pressure contraction \cite{Notari:2002yc, Finelli:2002we, Tsujikawa:2002qc}, or during a period of matter-dominated pressure-free contraction \cite{Wands:1998yp, Finelli:2001sr}. In all of these scenarios, expanding and contracting, if more than one field is present then in addition to density/curvature perturbations there can occur an amplification of entropy/isocurvature perturbations. What is more, entropy perturbations can act as a source for curvature perturbations, and in this manner the curvature perturbations present at the onset of the hot big bang phase can have an involved pre-history. (Note in this context that once/if the universe reaches thermal equilibrium during the hot big bang phase, the entropy perturbations disappear \cite{Weinberg:2004kf}, which implies that it is not surprising that none have been seen to date in the CMB data \cite{Ade:2015xua}.) 

The vast majority of cosmological models studied to date have assumed that the curvature perturbations of interest have either been fully created only during an expanding or a contracting phase of cosmological evolution. In the present paper, we want to explore a different possibility: namely that entropy perturbations are created during a contracting phase of the universe, and that they act as an important source for the curvature perturbations only at the beginning of the current expanding phase. We show that such a scenario offers interesting new possibilities, due to a possible non-trivial evolution of entropy perturbations during the bounce phase joining the contracting and expanding phases together. In particular, we will show with the example of a specific bounce model that entropy perturbations can grow significantly during the bounce phase, with the consequence that these enhanced entropy perturbations can easily provide the dominant contribution to the final post-bounce amplitude of the curvature perturbations.

We will consider two types of models: ekpyrotic models in which essentially only entropy perturbations are amplified during the contraction phase, and matter-dominated contraction models in which both curvature and entropy perturbations are created in roughly equal measure during the contraction phase. We then follow these perturbations across a non-singular ghost condensate bounce, where the long-wavelength curvature perturbations of interest remain conserved, while the entropic fluctuations are amplified further. In the ensuing expansion phase the entropy perturbations are then converted into curvature fluctuations. This has several broad consequences:

\begin{itemize}
\item The amplitude of the final curvature perturbations tends to be significantly enhanced. Conversely, this means that the entropy perturbations at the end of the ekpyrotic or matter-dominated contracting phases can be smaller than currently assumed in these models. In other words, the energy scale of the contraction phase can be smaller.
\item The enhancement of the amplitude of the entropy perturbations implies that the importance of its intrinsic non-Gaussianity is reduced. This is because the most important contribution to the non-Gaussianity of the final curvature perturbation comes from the non-linearity of the conversion process, and not from the intrinsic non-Gaussianity that can already be present in the entropy perturbations.
\item In matter-dominated contraction phases primordial gravitational waves are also produced, with an amplitude that is comparable to that of the curvature perturbations that are produced simultaneously \cite{Wands:1998yp}. This leads to a tension with current observations. In the models that we study here, the entropy perturbations grow to become even larger than the curvature perturbations produced during contraction, and the final amplitude of curvature perturbations (after conversion during the expanding phase) can thus be significantly enhanced compared to the tensor perturbations with the consequence that the tensor-to-scalar ratio is typically reduced by one or more orders of magnitude, in some cases bringing it close to current observational bounds. 
\end{itemize} 

We will begin with a review of non-singular ghost condensate bounce models in Sec.~\ref{section:bounce}, where we also discuss the evolution of both curvature and entropy perturbations in bouncing space-times. These results can then be used in specific ekpyrotic (Sec.~\ref{section:ekpyrotic}) or matter-dominated models (Sec.~\ref{section:matter}). We end with a discussion in Sec.~\ref{section:discussion}.

%%%%%%%%%%%%%%%%%%%%%%%%%%%%%%%%%%%%%%%%%%%%%%%%%%%%%%%%%%

\section{Non-Singular Bounce Model}
\label{section:bounce}

We are interested in models of gravity minimally coupled to two scalar fields $\phi, \chi$ with an action of the form
\be \label{eq:actiongeneral}
S = \int \d^4 x \sqrt{-g} \left[\frac{R}{2} + P(X,\f) - \frac{1}{2}(\partial\chi)^2 - V(\phi,\chi)\right]\,,
\ee
where $X\equiv -\frac{1}{2} g^{\mu \nu }\p_{\mu} \phi \p_{\nu} \phi = - \frac{1}{2}(\partial\phi)^2$ and $V$ is a potential which we will describe in much more detail later. Here we take $\phi$ to be the field driving the bounce, while $\chi$ is assumed to be transverse (in field space) to the background trajectory during the bounce phase. In order to model a non-singular bounce we will allow $\phi$ to be governed by a kinetic term containing higher derivatives of the form 
\be
P(X, \f)= K(\f)X+Q(\f)X^2 \,.
\ee
During the ghost-condensate phase the function $K(\phi)$ changes its sign to negative values while the function $Q(\phi)$ is turned on; we will choose a specific form below. Such ghost condensate bounces have been discussed by several authors \cite{Creminelli:2006xe, Buchbinder:2007ad, Creminelli:2007aq, Lehners:2011kr, Cai:2012va}, and their properties are by now rather well understood: when $K$ changes sign the null energy condition can be violated, allowing for a bounce solution. Nevertheless the fluctuations do not become ghost-like, due to the influence of the higher-derivative terms \cite{ArkaniHamed:2003uy}. In fact, detailed studies of the perturbative \cite{Koehn:2015vvy} and semi-classical stability \cite{Krotov:2004if} properties have been completed, and one can even formulate such models in ${\cal N}=1$ supergravity \cite{Koehn:2013upa}. 

In this paper we will consider bounces that are caused by the ghost-condensate phase of the field $\phi$, which violates the null energy condition.  Cosmological bounces can also be caused by quantum gravity effects, for example in loop quantum cosmology \cite{Ashtekar:2006wn}.  Interestingly, at least in the case that the bounce is dominated by an ekpyrotic-like field, the qualitative predictions concerning the scalar and tensor power spectra do not appear to depend significantly on the exact mechanism of the bounce \cite{Wilson-Ewing:2013bla, Cai:2014zga} and therefore the qualitative results obtained in this paper may also hold in other cases where the bounce occurs due to different physical effects, including quantum gravity effects.

\subsection{Background Evolution}

In a flat Robertson-Walker background with the metric
\be
\mathrm{d}s^2 = - \mathrm{d}t^2 + a(t)^2 \delta_{ij} \mathrm{d}x^i \mathrm{d}x^j \,,
\ee
the background equations of motion are 
\bea
3H^2&=& \frac{1}{2}K(\f) \dot \f^2 + \frac{3}{4} Q(\phi) \dot\phi^4 + \frac{1}{2} \dot\chi^2+V(\f,\chi)\\
\dot H&=& - \frac{1}{2} K(\f) \dot \f^2 - \frac{1}{2} Q(\phi) \dot\phi^4 - \frac{1}{2} \dot \chi^2 \label{eq:Hdot}\\
0 &=& P_{,\f} - V_{,\phi} -P_{,X}(\ddot \f+3H \dot \f)-P_{,XX}\ddot \f\dot \f^2-P_{,X\f}\dot \f^2\\
0&=& \ddot \chi+3H \dot \chi+V_{,\chi} \label{eq:P}
\eea
During the bounce phase, $\chi$ is taken to be constant, and will thus not contribute to the dynamics of the background (but we will be highly interested in the perturbations of $\chi$). In the kinetic term for $\phi$ we will use the following form of the functions $K$ and $Q,$  \cite{Koehn:2013upa}
\bea
K(\f)&=& 1-\frac{2}{\left (1 +\frac{1}{2} \f^2 \right)^2 }\,,\\
Q(\f)&=&\frac{q}{\left (1 +\frac{1}{2} \f^2 \right)^2 }\,,
\eea
while we take the potential to be given by a regularised version of an ekpyrotic potential \cite{Koehn:2015vvy}
\bea \label{eq:potential}
V(\f)&=&-\frac{2V_o}{e^{-\sqrt{2 \ep} \f }+e^{\sqrt{2 \ep} \f }}\,,
\eea
where $V_o$ and $\epsilon$ are constants. Away from the bounce, there will be additional contributions to the potential, which depend on the specific model under consideration and which we will discuss at the relevant locations.

\begin{figure}[htbp]
	\begin{minipage}{0.5\textwidth}
		\includegraphics[width=0.92\textwidth]{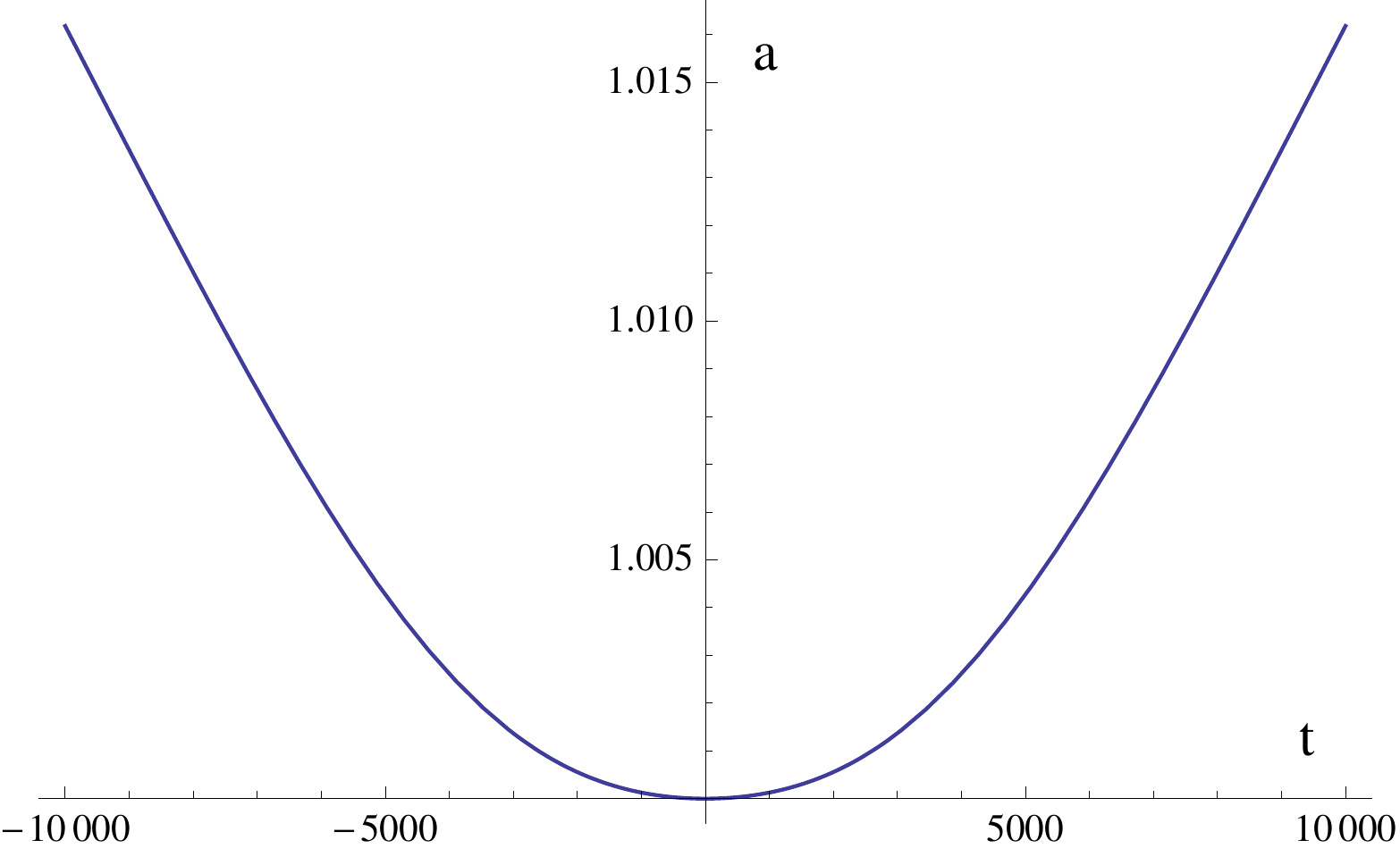}
	\end{minipage}%
	\begin{minipage}{0.5\textwidth}
		\includegraphics[width=0.92\textwidth]{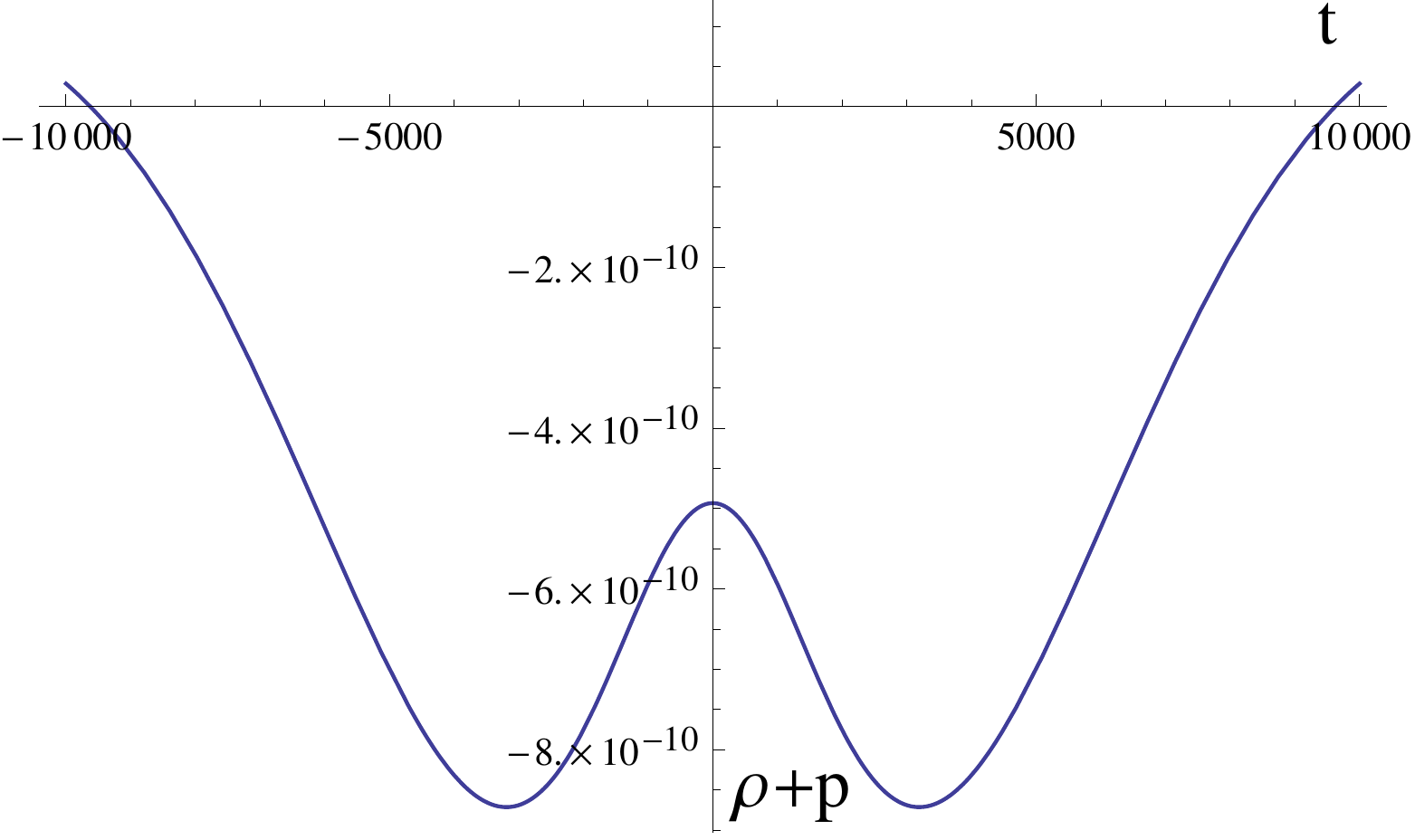}
	\end{minipage}%
	\caption{\footnotesize \hangindent=10pt
A numerical solution around the time of the non-singular bounce. The left panel shows the evolution of the scale factor $a$, while the right panel plots the sum of energy density $\rho$ and pressure $p$. When this last sum is negative, the null energy condition is violated, which is a necessary condition in order to obtain a non-singular bounce in a flat Robertson-Walker universe.} 
	\label{fig:a}
\end{figure}

A numerical solution of the equations of motion near the bounce is shown in Fig.~\ref{fig:a}. Here the parameters of the model are chosen to be
\be \label{eq:parameters}
\ep=10 \, , \qquad V_o=2 \times 10^{-8} \, , \qquad q=10^8 \, ,\qquad 
\ee
where $q,$ with mass dimension of $-4,$ determines the energy scale of the bounce. In the present case the energy scale is $(10^{-2}M_{Pl})^4,$ two orders of magnitude below the reduced Planck mass. This scale also determines the duration of the bounce, i.e.~the time period over which the null energy condition is violated, $t \in (-\sqrt{q},\sqrt{q})=(-10^4,10^4),$ as is verified in the right panel of Fig.~\ref{fig:a} which plots the sum of energy density and pressure. The initial conditions for the bounce solution were chosen such that the bounce occurs at $t=0$ with 
\be
\phi(0)=0 \, ,\qquad a(0)=1 \, ,\qquad
\chi(0)=0 \, ,\qquad \dot \chi(0)=0\,.
\ee
From the first Friedmann equation with $\rho=0$, we obtain $\dot \phi(0) \approx 9.7 \times 10^{-5},$ where we have chosen $\phi$ to roll from negative to positive values as time progresses.

Note that we have chosen $\phi(0) = 0$ at the bounce time since this is the centre of the region where the scalar field $\phi$ can violate the null energy condition and cause the bounce.  The bounce is symmetric around $t=0$ due to the symmetric form of $K(\phi)$ and $Q(\phi)$ (the form of the potential does not affect this since $\phi$ is kinetic-dominated during the bounce); we made these choices for the sake of simplicity.  An asymmetric bounce is also possible (see, e.g., \cite{Battarra:2014tga}), but the evolution of the long-wavelength perturbations (those of observational interest) is not significantly affected by the details of the bounce since the bounce time is much shorter than the wavelength of these perturbations.

One may also approximate the numerical solution analytically near the bounce. Note that at the bounce itself the Hubble rate is zero by definition, $H=0,$ and thus the Friedmann equation implies that the energy density vanishes too. Combining this result with the equation of motion for the scale factor \eqref{eq:Hdot} and neglecting the potential at the bounce then gives
\bea \label{a-bounce}
a(t)&\approx& e^{t^2/(18 q)}\,,\\
\phi(t)&\approx& \sqrt{\frac{2}{3q}} \, t\,.
\eea
The scalar field $\phi$ evolves linearly in time near the bounce, which is the characteristic time evolution of a ghost condensate. 

The bounce is followed (and, depending on the model, perhaps also preceded) by a phase of kinetic energy domination, during which the kinetic term for $\phi$ is to a good approximation canonical, and with equation of state $w=1$. The background equations then reduce to the simple form
\be
3H^2\approx\frac{1}{2} \dot \phi^2\approx-\dot H, \qquad
\ddot \phi+3H \dot \phi\approx0\,,
\ee
and the solution is given by
\be
a\approx \bar{a}_o (\pm t)^{1/3}, \qquad \phi \approx \sqrt{\frac{2}{3}} \ln(\pm t)+\bar{\phi}_o,
\ee
for some constants $\bar{a}_o,\bar{\phi}_o,$ and where a positive (negative) sign corresponds to an expanding (contracting) kinetic phase. The energy density scales as $\rho \propto a^{-6}$, since
\be
\dot \phi \approx \sqrt{\frac{2}{3}} \frac{\bar{a}^3_o}{a^3} \, .
\ee
We are now ready to turn our attention to the behaviour of fluctuations in this background.

%%%%%%%%%%%%%%%%%%%%%%%%%%%%%%%%%%%%%%%%%%%%%%%%
\subsection{Adiabatic and Entropy Perturbations Through the Bounce}

An important simplification of the background we are studying is that only one of the scalar fields, namely $\phi,$ evolves at the background level. This has the consequence that during the bounce phase, the perturbations of $\phi$ (which in their gauge-invariant form correspond to curvature perturbations $\cal R$) and the perturbations of $\chi$ (which correspond to entropy perturbations) evolve independently, and we can study them separately. 

\begin{figure} 
	\centering
	\includegraphics[width=0.5\textwidth]{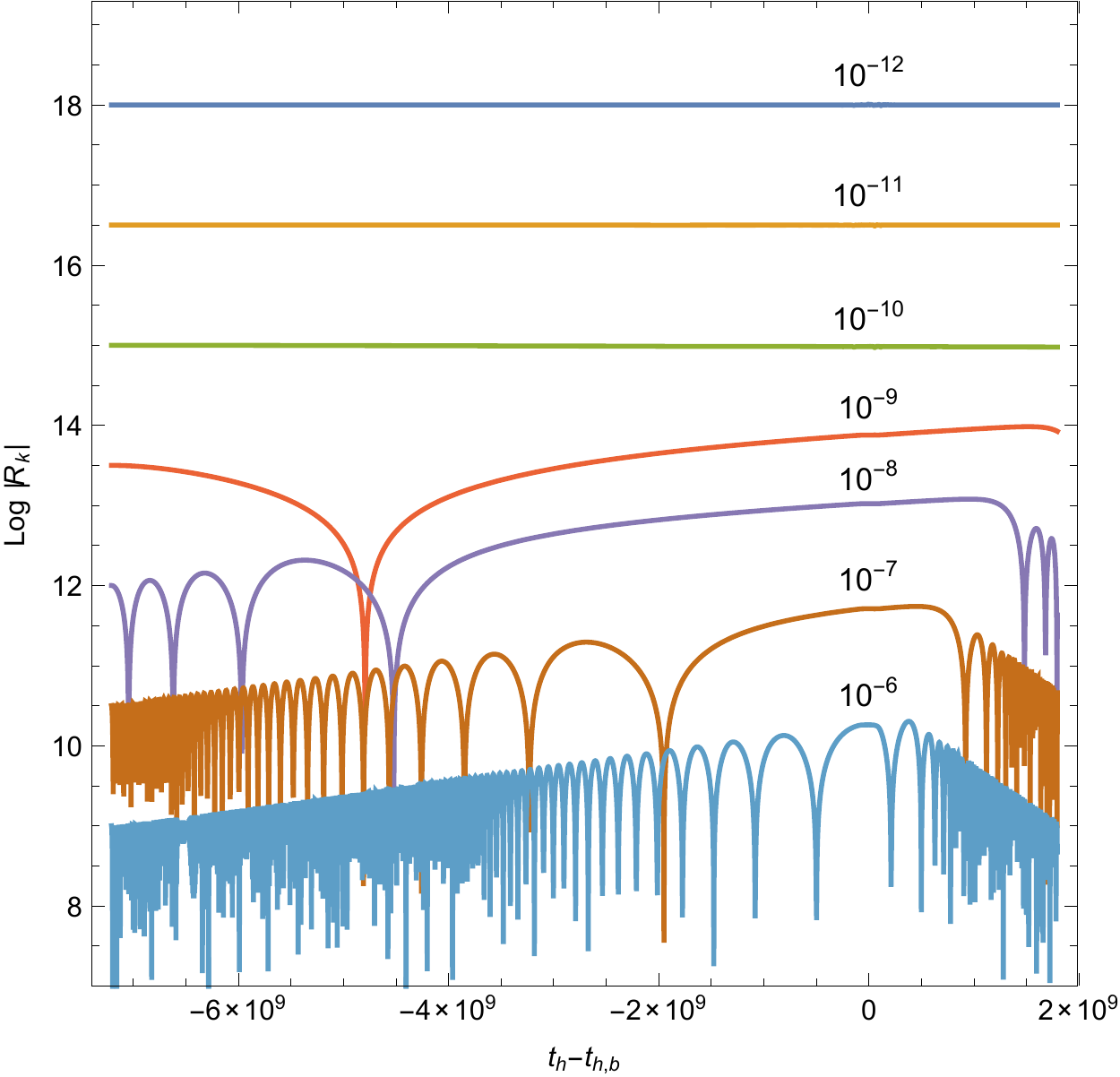}
	\caption{\footnotesize \hangindent=10pt
Evolution of the curvature perturbations (here denoted by ${\cal R}_k$) across a flat non-singular ghost condensate bounce, for different wavenumbers $k$ (this figure is taken from \cite{Koehn:2015vvy}). We are interested in long-wavelength modes which are conserved with high accuracy. The numerical plot here shows the evolution of the curvature perturbation modes in terms of harmonic time $t_h,$ defined as $\mathrm{d}t_h = a^2 \mathrm{d}\tau = a^3\mathrm{d}t.$ The time interval plotted above is much longer than the period of violation of the null energy condition; see \cite{Battarra:2014tga,Koehn:2015vvy} for more details.}
	\protect
	\label{fig:Modes}
\end{figure}

The evolution of the curvature perturbations has been studied in detail in \cite{Battarra:2014tga}, here we will simply summarise the main results. In conformal time, the linearised equation of motion for the comoving curvature perturbation ${\cal R}$ (which on large scales can be thought of as a local perturbation in the scale factor of the universe) is given by
\begin{equation} \label{eq:conformalR2}
\frac{ d ^2 \mathcal{R}}{d \tau ^2} + \frac{2}{z} \frac{ d z }{ d \tau} \frac{ d \mathcal{R}}{ d \tau} + c  _{s} ^2 k ^2 \mathcal{R} = 0 \;,
\end{equation}
where 
\begin{eqnarray}
z ^2 & = & a^2 \frac{P_{,X} X + 2 P_{,XX}X^2}{H^2}  \;,\label{eq:z22} \\
c_s ^2 & = &  \frac{P_{,X}}{ P_{,X} + 2 P_{,XX} X} \;.\label{eq:cs22}
\end{eqnarray}
Then, near the bounce (taken to occur at $\tau=0$) this equation is solved by the series
\begin{equation}
\mathcal{ R} = \alpha  \left(1 - \frac{1}{2} \bar{c}_s ^2 k ^2 \tau^2 + \ldots \right) + \beta \tau^3 \;,
\end{equation}
where $\alpha,\beta$ are constants. Here the value of the speed of sound at the bounce is approximated by $\bar{c}_s^2 \approx -1/3.$ Thus, for perturbations which have a wavelength that is long compared to the scale of the bounce ($k/a \ll q^{1/4}$), we expect the solution to be approximately constant across the bounce. This is confirmed by numerically solving the equation of motion, see Fig.~\ref{fig:Modes}. In the figure, the purple and red lines correspond approximately to the scale of the bounce, and here some evolution of the curvature perturbation is seen. However, for longer-wavelength perturbations (all curves that are above the red line) there is essentially no evolution at the bounce. Modes of potential interest for cosmological observations have a wavelength many orders of magnitude larger still, and such modes remain conserved across the bounce to very high precision.  (As an aside, note that the Mukhanov-Sasaki perturbation variable $v = z \mR$ blows up across the bounce, as noticed in \cite{Cai:2013kja}, but this is due to $z$ diverging at the bounce point.  The perturbation of physical interest, $\mR$, remains frozen on large scales during the bounce.)

We should add one comment regarding the adiabatic modes: as one can see both from the series solution above and the figure, short-wavelength modes are affected by the gradient instability that is caused by the speed of sound squared going negative during the bounce phase. This instability is in fact ever stronger for ever shorter modes. It is thus of importance to know the cut-off of our effective field theory, in order to assess whether the background bounce solution is really trustworthy or not. The cut-off scale $\Lambda_{strong}$ was calculated recently in \cite{Koehn:2015vvy} and it was found that it must be evaluated at the moment when $P_{,X}=0,$ with the result that 
\be
\Lambda_{strong}^4 = \frac{2}{q}\,, \qquad \rho\mid_{P_{,X}=0} = \frac{1}{4q} + V(\phi,\chi)\,,
\ee
where we have also listed the energy scale of the background for comparison. As one can see, in the absence of a potential the cut-off is a factor of $8$ above the energy density of the background, and this separation of scales is larger when a negative potential is present, as we have assumed here (in our specific example these two scales are separated by a factor of about $40$). Meanwhile, the cut-off is not too far above the energy scale of the background, implying that the most dangerous very short/high frequency perturbation modes indeed lie outside the regime of validity of the theory. Thus, from an effective field theory point of view, this bounce model is reliable%
\footnote{Note that ghost condensate (and also Galileon) bounces necessarily contain a period during which the speed of sound squared is negative, though this period may occur before, during or after the bounce \cite{Easson:2011zy,Easson:2013bda,Libanov:2016kfc,Kobayashi:2016xpl}, and it can be shifted to lie outside of the period where the null energy condition is violated \cite{Ijjas:2016tpn}. The important point is that all fluctuations which are within the regime of validity of the (effective) theory remain under control, which for the model used here was shown to be the case in \cite{Koehn:2015vvy}.}.
The presence of a negative potential during the bounce phase then suggests that the conversion of entropy into curvature perturbations ought to occur after the bounce, and not before \cite{Koehn:2015vvy}, which provides additional motivation for the present study.

Now we can start to analyse the evolution of the entropy perturbations, which is the main focus of this paper. We should introduce some notation first. The formalism suggesting a decomposition of perturbations into adiabatic and entropy modes was first discussed in \cite{Gordon:2000hv} at the linear level, and later extended by various authors \cite{GrootNibbelink:2000vx,Rigopoulos:2005xx,Langlois:2006vv}. The idea is to introduce a direction in field space tangent to the background trajectory (denoted $\sigma$ and representing the adiabatic direction) and one perpendicular to it (denoted $s$ and representing the entropy direction). If we group our two scalars together as $\phi^I = (\phi, \chi)$ then we can define 
\be
e_\sigma^I \equiv  \frac{1}{\sqrt{\dot\phi_1^2+\dot\phi_2^2}}
\left(\dot \phi_1, \dot \phi_2 \right),
\qquad e_s^I \equiv
 \frac{1}{\sqrt{\dot\phi_1^2+\dot\phi_2^2}}
\left(- \dot \phi_2, \dot \phi_1 \right)
 \label{e1} .
\ee
It is also convenient to introduce the angle in field space $\theta$ defined by
\be
\cos \theta \equiv \frac{\dot\phi_1}{\sqrt{\dot\phi_1^2+\dot\phi_2^2}}\, ,
\qquad \sin \theta \equiv
\frac{\dot\phi_2}{\sqrt{\dot\phi_1^2+\dot\phi_2^2}}\, ,
\label{theta}
\ee
so that
\be
e_\sigma^I =
\left(\cos \theta,\sin \theta  \right),
\qquad e_s^I =
\left(-\sin \theta , \cos \theta \right),
\ee
and the notation 
\be
\dot{\sigma} \equiv  \sqrt{\dot\phi_1^2+\dot\phi_2^2}\,.
\ee
At linear and second order, the adiabatic and entropy perturbations are then defined to be 
\be 
\delta \sigma^{(1)} = e_{\sigma I} \delta \phi^I\,, \qquad \delta s^{(1)} = e_{sI} \delta \phi^I\,,
\ee
\be
\delta \sigma^{(2)} = e_{\sigma I} \delta \phi^{I(2)} + \frac{1}{\dot\sigma}\delta s \dot{\delta s}\,, \qquad \delta s^{(2)} = e_{sI} \delta \phi^{I(2)} - \frac{\delta \sigma}{\dot\sigma}\left( \dot{\delta s} + \frac{\dot\theta}{2}\delta \sigma \right)\,,
\ee
where we do not specify the order in perturbation theory when no confusion may arise. Note that the adiabatic perturbations are not gauge-invariant; rather, the gauge-invariant quantity associated with them is nothing but the curvature perturbations discussed above (at linear order ${\mathcal{R}} = \psi - (H/\dot\sigma) \delta \sigma,$ where $\delta g_{ij} = 2a^2 \delta_{ij} \psi$).  On the other hand, the entropy perturbations are gauge-invariant by construction. Up to second order and on large scales, they obey the equation of motion
\bea \label{eq:oemds}
\ddot{\de s} +3 H \dot{\de s}+ \left( V_{ss}+3  \dot \theta^2  \right) \de s&=&\\ 
-\frac{ \dot \theta }{\dot \sigma}  (\dot{\de s})^2  -\frac{2}{\dot \sigma}  \left( \ddot  \theta+V_{\sigma}\frac{ \dot \theta}{\dot \sigma} -\frac{3}{2}H \dot \theta \right ) \de s \dot{\de s}  &+& \left (-\frac{1}{2} V_{sss}+ 5V_{ss}\frac{\dot \theta}{\dot \sigma}  +9\frac{ \dot \theta^3}{\dot \sigma}  \right) (\de s)^2, \nonumber
\eea
where $V_{\s} \equiv e_{\s}^I V_{,I}$, $V_{s} \equiv e_{s}^I V_{,I}$, $V_{ss} \equiv e_{s}^I e_{s}^J V_{,IJ},$ etc.~(a useful additional relation is $V_{s} = - \dot\sigma \dot\theta$), and where $\delta s = \delta s^{(1)} + \delta s^{(2)}.$ Also, at large scales the spatial gradient terms are negligible and have been dropped in the above expression.

During the bounce phase, the field space trajectory is straight ($\dot\theta=0$), and the equation thus simplifies to 
\bea \label{eq:oemds2}
\ddot{\de s} +3 H \dot{\de s}+ V_{ss}\de s + \frac{1}{2} V_{sss} (\de s)^2 = 0\,.
\eea

\begin{figure}[htbp]
	\begin{minipage}{0.5\textwidth}
		\includegraphics[width=0.92\textwidth]{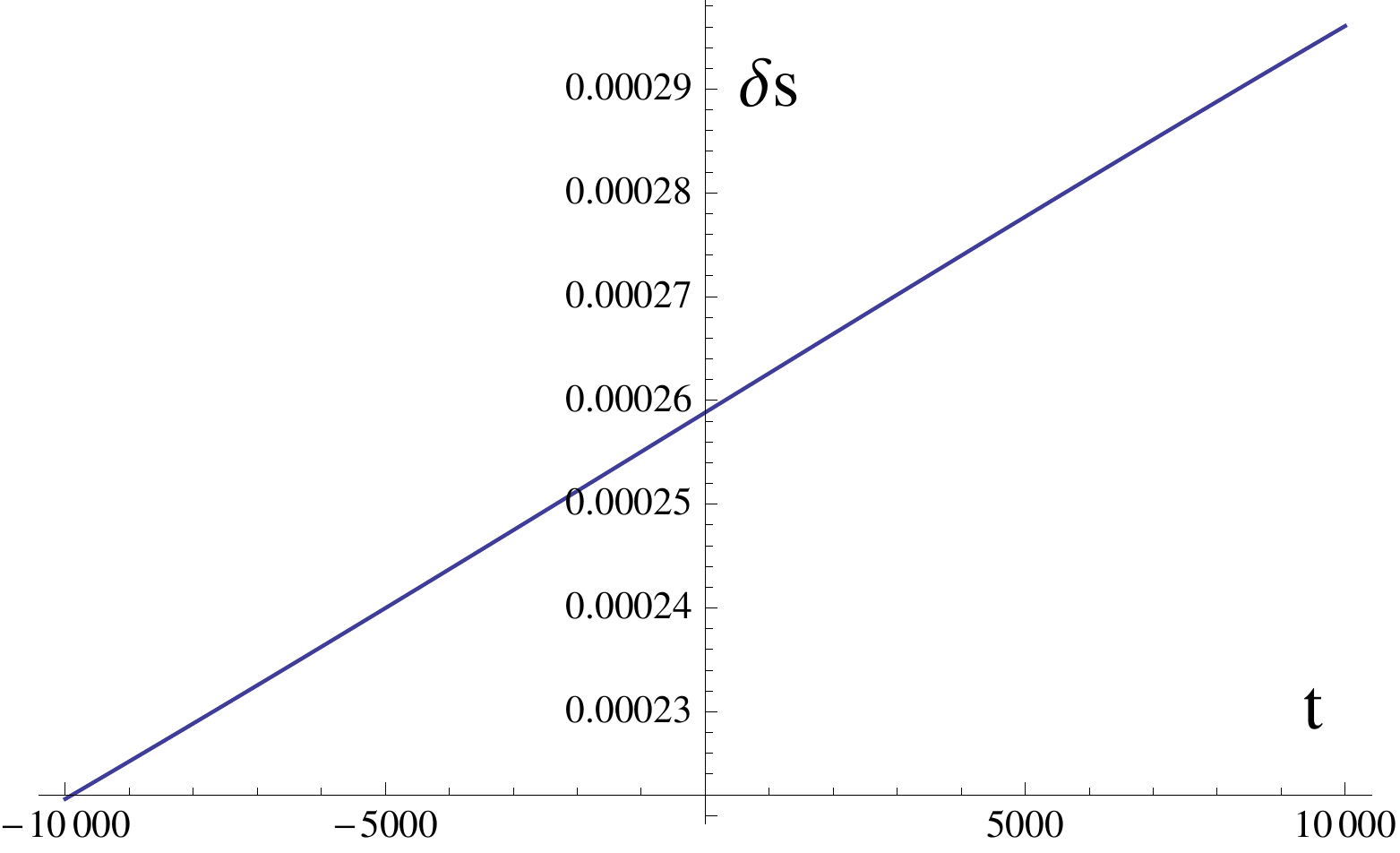}
	\end{minipage}%
	\begin{minipage}{0.5\textwidth}
		\includegraphics[width=0.92\textwidth]{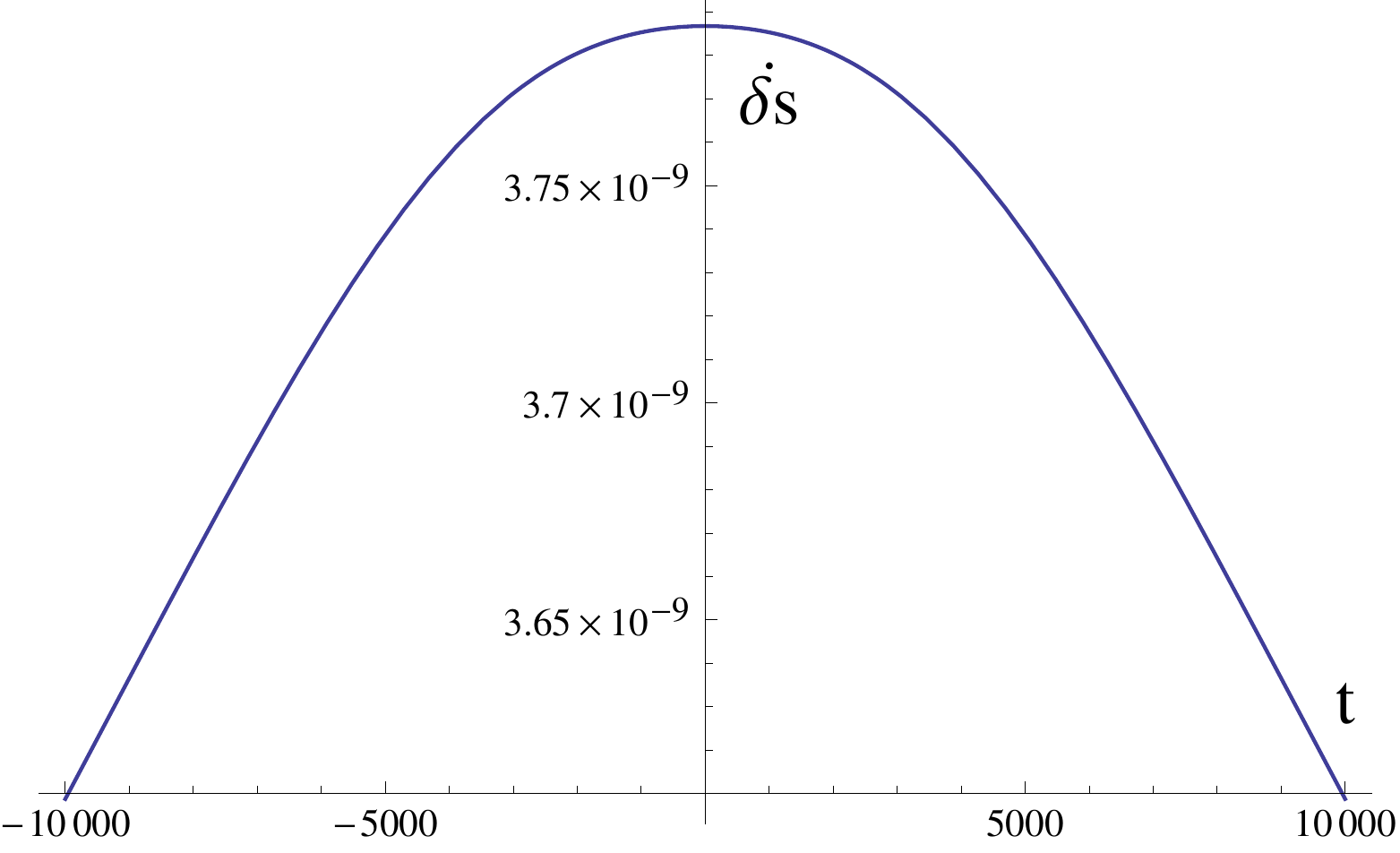}
	\end{minipage}%
	\caption{\footnotesize \hangindent=10pt
A plot of the entropy perturbation and its time derivative during the period where the null energy conditions is violated, and where the bounce occurs. The parameters are those of Eq.~\eqref{eq:parameters}. In this example, the transverse potential (i.e.~the potential in the entropic direction) is flat, and there is a modest growth of the perturbation across the bounce.} 
	\label{fig:s_stable}
\end{figure}

Let us first analyse the case where the transverse potential is flat, $V_{ss} = V_{sss} = 0$. In that limit, $\dot{\de s} = A \, a(t)^{-3}$, which given \eqref{a-bounce} is easily integrated,
\be
\de s = A \, \sqrt \fr{3 \pi q}{2} \, \mathrm{Erf} \left( \fr{t}{\sqrt{6 q}} \right) + B,
\ee
where $A$ and $B$ are integration constants determined by the values of $\de s$ and $\dot{\de s}$ at the beginning of the bounce phase.  During the bounce period, the entropy perturbation thus changes by an amount
\be \label{deltadeltas}
\Delta \delta s  = A \int _{-t_b=-q^{1/2}}^{t_b=q^{1/2}}e^{-\frac{1}{6q}t^2} dt = A \, \sqrt{6\pi q} \, \mathrm{Erf}(1/\sqrt{6}) \approx 1.9 \times A \sqrt{q}\,.
\ee
In cosmological models such as those that we consider, the entropy perturbation before the bounce typically reaches a scale not much smaller than the scale of the bounce, i.e.~one would expect that the amplitude of the entropy perturbations is of the order of $q^{-1/2}$, and that therefore $A q^{1/2} \sim {\cal O}(1)$ (this estimate will be justified in Sec.~\ref{section:ekpyrotic}).  In this case there is a modest growth of the entropy perturbation during the period of NEC violation. A numerical example confirming this estimate is provided in Fig.~\ref{fig:s_stable}, where the initial entropy perturbation was taken to be $\delta s_{pre-bounce} = 10^{-5}.$  The small growth of the entropy perturbations across a non-singular bounce was previously noted in \cite{Cai:2013kja}.

\begin{figure}[htbp]
	\begin{minipage}{0.5\textwidth}
		\includegraphics[width=0.92\textwidth]{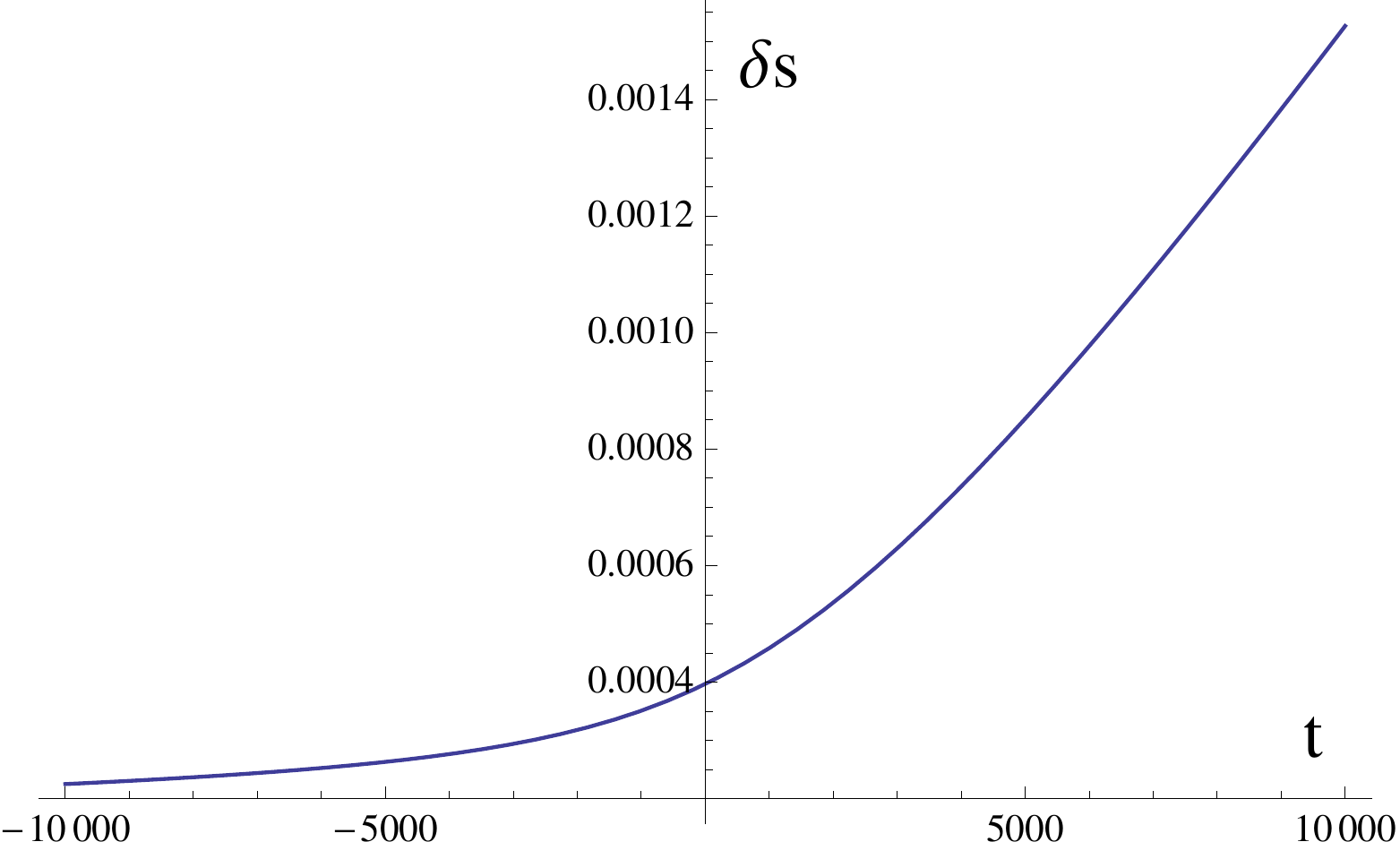}
	\end{minipage}%
	\begin{minipage}{0.5\textwidth}
		\includegraphics[width=0.92\textwidth]{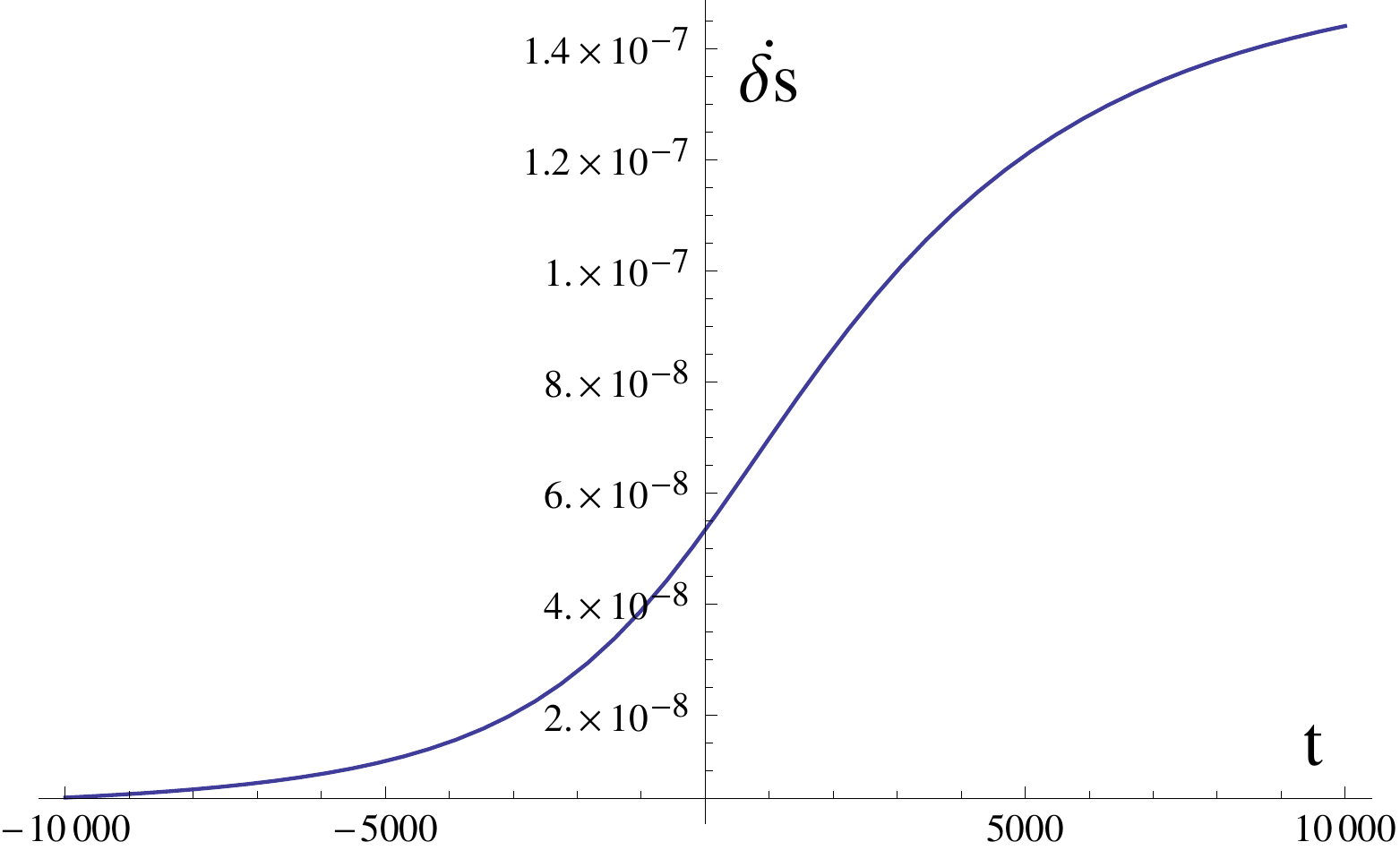}
	\end{minipage}%
	\caption{\footnotesize \hangindent=10pt
A plot of the entropy perturbation and its time derivative during the period where the null energy conditions is violated and the bounce occurs. The parameters are again those of Eq.~\eqref{eq:parameters}, but in this example the transverse potential is unstable, see Eq.~\eqref{eq:unstable}. The instability causes an additional growth in the entropy perturbation across the bounce.} 
	\label{fig:s_unstable}
\end{figure}

An important point here is that in the case that $V_{ss} = V_{sss} = 0$ during the bounce, for this amplification to occur it is essential that the entropy perturbations never freeze. If the entropy perturbations are frozen ---i.e.~if $\dot{\de s} = 0$--- at the onset of the bounce phase, then the constant solution $\de s = B$ is singled out and no amplification will occur during the bounce.  So, a necessary condition for the entropy perturbations to be amplified during the bounce is that their time evolution remain non-trivial at all times leading up to the bounce.

We can now also look at the case where a transverse, unstable potential may be present during the bounce phase. Such potentials are motivated by two-field ekpyrotic models, as we will discuss in more detail in the next section. We take the potential to be given by the form 
\be \label{eq:unstable}
V(\f)=-\frac{2V_o}{e^{-\sqrt{2\ep} \f }+e^{\sqrt{2 \ep} \f }} \left( 1+ \frac{1}{2} \ep \c^2+\cdots  \right)\,,
\ee
implying that
\be
V_{ss}=-\frac{2V_o\ep}{e^{-\sqrt{2\ep} \f }+e^{\sqrt{2 \ep} \f }}\,.
\ee
During the bounce $\phi$ grows linearly in time, $\phi=\sqrt{2/(3q)} \, t$ leading to
\be
V_{ss}=-\frac{2V_o\ep}{e^{-\sqrt{4\epsilon/(3q)}  t }+e^{\sqrt{4\epsilon/(3q)}  t }}=-\frac{V_o\ep }{\cosh(\sqrt{4\epsilon/(3q)}  t)}
\ee
This unstable potential leads to an additional amplification of the entropy perturbations.  A typical example is shown in Fig.~\ref{fig:s_unstable}, where we have chosen the same parameters as in Eq.~\eqref{eq:parameters}. In this example, there is a total growth by a factor of about $7$ across the bounce. The growth does depend on the parameters of the model, in particular on $\epsilon$ which sets the scale of the instability. However, provided that the overall scale of the potential changes in proportion to the change of scale of the bounce, i.e.~provided $V_o \propto q^{-1},$ then the total growth is independent of the bounce energy scale $q^{-1/4}.$  Note that this additional growth in the amplitude of the entropy perturbations will be scale-invariant (i.e.\ independent of the Fourier wavenumber) since all of the Fourier modes of observational interest are in a regime where the gradient terms in their equations of motion are entirely negligible during the bounce.  Hence an unstable transverse potential leads to an additional growth, but at the expense of requiring more special initial conditions since now only trajectories lying very close to the ridge in the potential will make it through the bounce.  We will return to this point in Sec.~\ref{section:ekpyrotic}.

\begin{figure}[htbp]
	\begin{minipage}{0.5\textwidth}
		\includegraphics[width=0.92\textwidth]{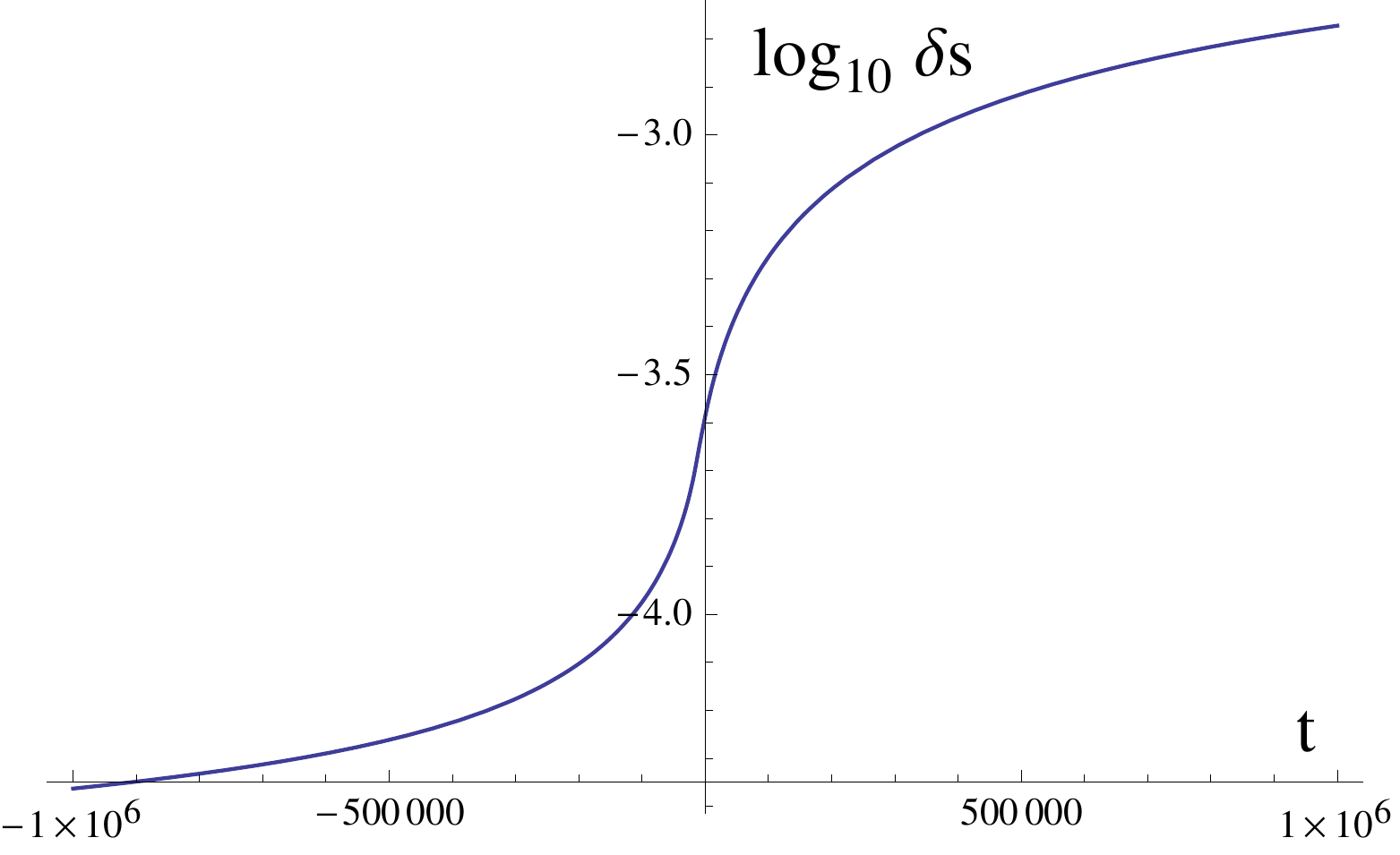}
	\end{minipage}%
	\begin{minipage}{0.5\textwidth}
		\includegraphics[width=0.92\textwidth]{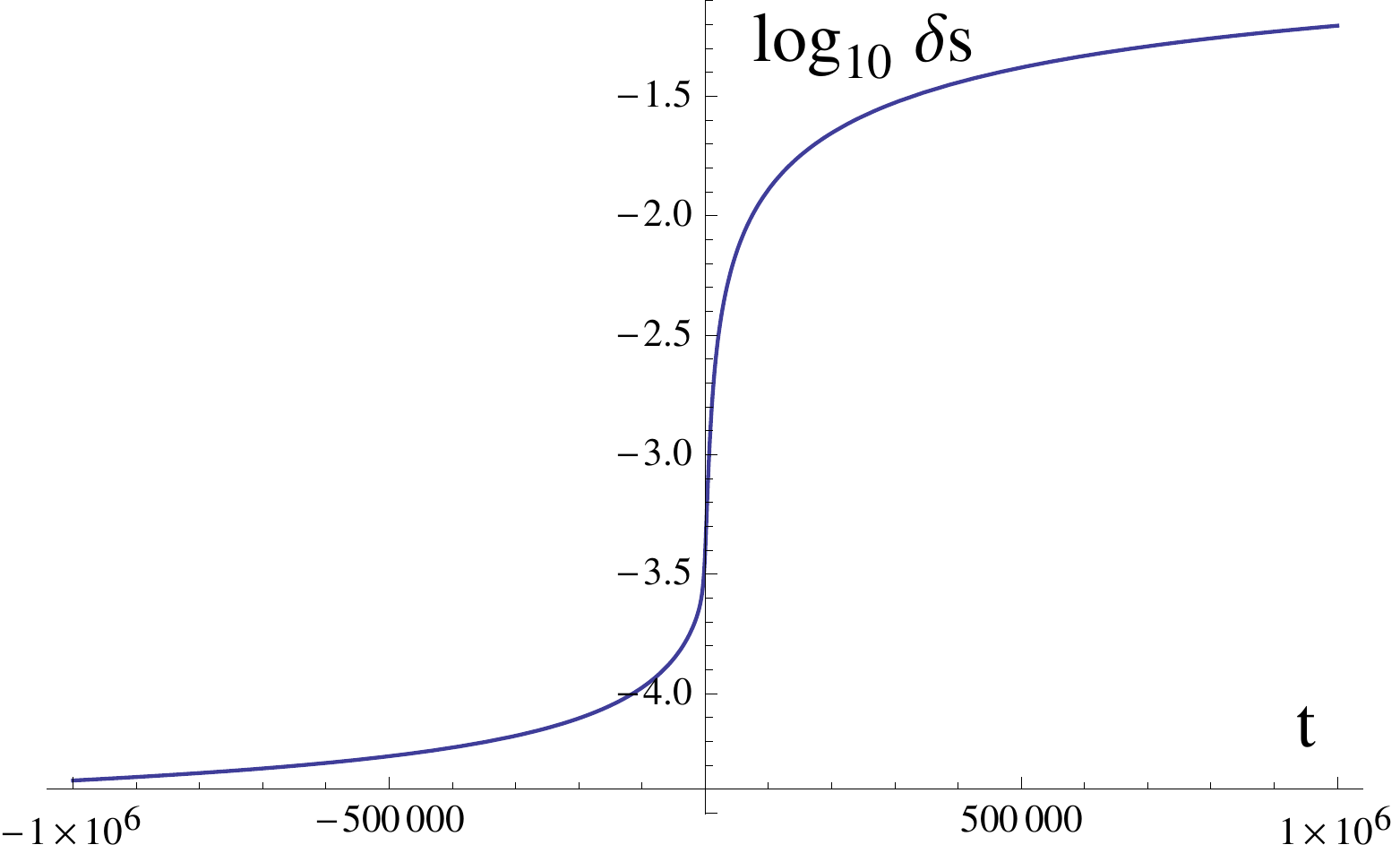}
	\end{minipage}%
	\caption{\footnotesize \hangindent=10pt
The growth of the entropy perturbations across the kinetic and bounce phases (where the period of null energy violation is confined to the interval $-10^4 < t < 10^4$). The left panel shows the case where the potential is flat in the entropic direction, while in the right panel this transverse potential is unstable. The combined effect of the kinetic and bounce phases leads to a significant overall amplification of the entropy perturbations.} 
	\label{fig:s_kin_bounce}
\end{figure}

The bounce is followed (and, if the potential is less important just before the bounce, also preceded) by a phase of kinetic energy domination. During these phases $H=1/(3t)$ and the potential is subdominant, such that the entropy perturbation obeys the approximate equation $\ddot{\de s} + (1/t)\times \dot{\de s} \approx 0$ and consequently grows approximately logarithmically,
\be
\de s = C \ln (t/t_k) + D
\ee
for some integration constants $C,D,t_k.$  Once again, this growth is conditional upon the entropy perturbations not being frozen, i.e.~that $\dot{\de s} \neq 0$ --- otherwise $\de s = D$ and no growth will occur.  Even for $C \neq 0$, the growth is rather slow, but the cumulative effect over the two (bounce--expanding kinetic) or three (contracting kinetic--bounce--expanding kinetic) phases can be substantial. We have illustrated this in Fig.~\ref{fig:s_kin_bounce}, where the left panel corresponds to having a flat transverse potential and the right panel shows the case of an unstable transverse potential. Even without an instability, the entropy perturbation tends to grow by $1.5$ to $2$ orders of magnitude, depending on how long the kinetic phases last. As we will see in detail in the following sections, this growth implies that a conversion of entropy perturbations into curvature perturbations after the bounce phase easily provides the dominant contribution to the final amplitude of the curvature perturbations, keeping in mind that any curvature perturbations that might be present before the bounce remain constant across that phase. The right panel in the figure then shows the amplification of the entropy perturbations for the case where the potential in the entropic direction is unstable. In that case the growth is further enhanced, and in our numerical example the entropy perturbation reaches an overall growth of $3$ orders of magnitude. Note that the bounce has the effect of not just amplifying the entropy perturbation, but also its time derivative $\dot{\delta s}$ grows during the bounce, thus bringing with it an even stronger growth during the expanding kinetic phase.

The next sections will explore the consequences of this growth of the entropy perturbations across the kinetic and bounce phases for specific cosmological models: ekpyrotic models in Sec.~\ref{section:ekpyrotic} and matter bounce models in Sec.~\ref{section:matter}.

%%%%%%%%%%%%%%%%%%%%%%%%%%%%%%%%%%%%%%%%%%%%%%%%%%%%%%%%%%
\section{Ekpyrotic Models}
\label{section:ekpyrotic}

Ekpyrotic models provide possible solutions to a number of early universe puzzles, such as the flatness problem, the horizon problem and the question of the origin of the primordial perturbations \cite{Khoury:2001wf, Lehners:2008vx}. They may also explain the classicality of the early universe \cite{Battarra:2014xoa, Battarra:2014kga, Lehners:2015sia}. We will briefly review the ekpyrotic solution to the flatness puzzle, as this allows us to introduce the relevant setting. We start with the Friedmann equation in the presence of different matter types, represented here by their energy densities $\rho_i,$ 
\be 
3 H^2 = \left(\frac{-3\kappa}{a^2}+ \frac{\rho_m}{a^3} +
\frac{\rho_r}{a^4}+ \frac{\sigma^2}{a^6} + \ldots +
\frac{\rho_{\phi}}{a^{2\epsilon}} \right) \label{Friedmann}
\ee 
The subscript $m$ refers to non-relativistic matter and includes dark matter, $r$ refers to radiation and $\sigma$ denotes the energy density of anisotropies in the curvature of the universe. In an expanding universe, as the scale factor $a$ grows, matter components with a slower fall-off of their energy density come to dominate. Thus, in an expanding universe the homogeneous curvature term $\frac{-3\kappa}{a^2}$ will be suppressed if there is a scalar field with equation of state $\epsilon < 1.$ This is the inflationary solution to the flatness problem. In a contracting universe, the situation is changed, as the homogeneous curvature term is suppressed in any case compared even to pressure-less matter. In that case, the dangerous term is the anisotropy term. But all of these terms are suppressed if there exists a scalar field with a very stiff equation of state $\epsilon > 3$ \cite{Erickson:2003zm}, called the ekpyrotic field. It can be modelled by a scalar field with a steep and negative potential (where $V_o$ is a constant),  
\be
V(\phi)=-V_o \, e^{{\sqrt{2\epsilon}} \phi},
\ee 
as we have effectively done in Eq.~\eqref{eq:potential}. (We assume that the ekpyrotic phase proceeds at negative values of $\phi$.) Then, during the ekpyrotic phase, the fields evolve according to the scaling solution
\be
a \propto (-t)^{1/\ep} \, , \qquad \phi = 
\sqrt{\frac{2}{\ep}} \ln{\left[ - \left( \frac{V_o \ep^2}{ (\ep -3)} \right)^{\frac{1}{2}}  t\right]} \,, \qquad \ep=\frac{c^2}{2} = \frac{3}{2} \left(1+ w \right). \label{scalingsolution}
\ee
If this phase lasts long enough, homogeneous curvature and anisotropies will be suppressed to such an extent that they become irrelevant for the rest of the evolution, which is what we will assume henceforth. 

We can now look at perturbations in this background. First, we should investigate the perturbations in the ekpyrotic scalar $\phi.$ As reviewed above, the gauge-invariant perturbation variable associated with $\phi$ is the curvature perturbation $\mathcal{R}.$ If we define a re-scaled variable $v=z\mathcal R,$ then its equation of motion is given by
\be
v'' + \left( k^2 - \frac{z''}{z} \right) v = 0,
\ee
where $z=a\sqrt{2\epsilon}$ during the ekpyrotic phase, and where a prime denotes a derivative with respect to conformal time $\tau$. For an exponential potential, $\epsilon$ is constant, and thus $z''/z = a''/a.$ Using $a \propto (-\tau)^{1/(\epsilon - 1)},$ we find 
\be
\frac{a''}{a} = - \frac{\ep - 2}{(\ep - 1)^2 \tau^2},
\ee
Imposing the usual boundary condition that in the far past and on small scales the mode functions are those of vacuum quantum fluctuations in Minkowski space, $\lim\limits_{k\tau \rightarrow -\infty}{v_s} = \frac{1}{\sqrt{2 k}} e^{- i k \tau},$ the solution is a Hankel function of the first kind with a prefactor independent of $k$,
\be
v = \frac{\sqrt{- \pi \tau}}{2} H_{\nu}^{(1)} (-k \tau) \qquad \text{with} \quad \nu = \frac{\ep - 3}{2(\ep - 1)},
\ee
up to an irrelevant global phase factor.
Given that $\ep > 3,$ we have $0 < \nu < 1/2.$ If we now inspect the late time behaviour of the mode functions,
\be
v \sim \frac{(-\tau)^{\frac{1}{2}-\nu}}{k^\nu}, \quad (|k\tau| \ll 1)
\ee
we can see that these modes are not amplified. Moreover, their spectral index, given by $n_s = 4-2\nu$ is highly blue, $3 < n_s < 4$ and thus these perturbations play no role at all on the large scales of observational interest \cite{Lyth:2001pf,Creminelli:2004jg}. In order to explain the temperature fluctuations of the cosmic microwave background, we must thus go beyond what we have so far. And in fact the situation changes significantly upon including a second scalar field $\chi.$

Two possibilities have been studied: adding a scalar $\chi$ with a canonical kinetic term and an unstable potential \cite{Notari:2002yc,Finelli:2002we,Lehners:2007ac}, or adding a scalar $\chi$ with a non-minimal kinetic coupling to $\phi$ and a stable potential \cite{Qiu:2013eoa,Li:2013hga,Fertig:2013kwa,Ijjas:2014fja}. In both cases, the perturbations in $\chi$ are entropy perturbations, which are amplified and can become (nearly) scale-invariant. These can then act as scale-invariant seeds for the curvature perturbations, so that one obtains the desired nearly scale-invariant primordial curvature perturbations via this two-step process. 

In the case where $\chi$ is added with a canonical kinetic term, the potential is taken to be of the parametric form
\be \label{eq:unstable3}
V(\f)=-\frac{2V_o}{e^{-\sqrt{2\ep} \f }+e^{\sqrt{2 \ep} \f }} \left( 1+ \frac{1}{2} \ep \c^2+\frac{1}{3!}\ep^{3/2} \kappa_3 \chi^3  \right)\,,
\ee
where $\kappa_3$ parameterises the tilt of the unstable direction. This form of the potential arises by considering two scalar fields with ekpyrotic potentials, and performing a rotation in field space \cite{Koyama:2007ag}.  At linear order, the equation of motion for $\de s$ is
\be 
\ddot{\delta s} + 3H\dot{\delta s} + \left(\frac{k^2}{a^2}
+ V_{ss} \right) \delta s = 0\,. \label{eq-entropylinear}
\ee
In conformal time, and for the re-scaled variable $v_s= a \de s,$ one obtains 
\be 
{v_s}'' + \left(k^2 -\frac{a''}{a} + a^2 V_{ss} \right) v_s = 0.
\label{eq-entropy-S}
\ee 
If one allows $\ep$ to vary slowly, for large $\ep$ this leads to the solution \cite{Lehners:2007ac}
\be 
v_s = \frac{\sqrt{-\pi\tau}}{2}H^{(1)}_\nu (-k\t) \qquad \text{with} \quad \nu = \frac{3}{2} \left(1 - \frac{2}{3 \epsilon} +\frac{\epsilon_{,{\cal N}}}{3 \epsilon} \right),
\ee 
where we have expressed the time variation of the equation of state in terms of the number of e-folds left before the end of the ekpyrotic phase ($d{\cal N} = -d \ln|aH|$). At the end of the ekpyrotic phase, the entropy perturbation is thus given by%
\footnote{Note that this calculation justifies the statement given below Eq.~\eqref{deltadeltas} concerning the typical amplitude of the entropy perturbations at the onset of the bounce phase.} 
\be 
\de s(t_{ek-end}) \approx \frac{|\epsilon V_{ek-end}|^{1/2}}{\sqrt{2}k^\nu}, \label{entropylineargenerated}
\ee
while the spectral index of the entropy perturbation is \cite{Lehners:2007ac} 
\be 
n_s = 1 + \frac{2}{\epsilon} -  \frac{d \ln \epsilon}{d {\cal N}}.
\ee
To estimate the spectral index, one may assume a scaling law $\epsilon \approx {\cal N}^{\alpha}$ \cite{Khoury:2003vb,Ijjas:2013sua}. Then, for $\alpha \approx 1$,  the spectral tilt is $n_s \approx 1+ 1/{{\cal N}} \approx 1.02$, and hence slightly blue.  However, in ekpyrotic models the steepness of the potential must decrease in order for the ekpyrotic phase to come to an end, and thus $\alpha$ should be a little larger. For $\alpha > 1.14$ the spectral tilt is red.  In particular, for $\alpha \approx 2$ one obtains values $n_s \approx 0.97$ in good agreement with Planck data \cite{Lehners:2013cka,Ade:2015xua}. At second order, and on large scales, the equation of motion is given by Eq.~\eqref{eq:oemds2}, which can be solved easily to yield (to leading order in $1/\epsilon$)
\be 
\delta s= \delta s^{(1)} + \frac{\kappa_3 \sqrt{\epsilon}}{8}\left(\delta s^{(1)}\right)^2 \,,
\label{entropyInitialCondition}
\ee 
where the linear, Gaussian part is given by $\delta s^{(1)} \propto 1/t.$ The second order correction parameterises the intrinsic (local) non-Gaussianity of the entropy perturbation, and it can be shown to depend on the tilt $\kappa_3$ of the transverse potential \eqref{eq:unstable3}. Let us point out here that the unstable transverse potential has the merit of amplifying the entropic perturbations. On the other hand, special initial conditions are required in order for the ekpyrotic phase to last long enough \cite{Buchbinder:2007tw}. This is generally seen as a drawback, but may also act as a kind of filter of initial conditions in a cyclic context, as in the phoenix picture of the cyclic universe \cite{Lehners:2008qe,Lehners:2009eg,Lehners:2011ig}.

There exists a second class of models which are stable and thus more robust in terms of initial conditions \cite{Qiu:2013eoa,Li:2013hga,Fertig:2013kwa,Ijjas:2014fja}. In these models, entropy perturbations may also be amplified, in this case due to a non-minimal coupling between the two scalar fields $\phi$ and $\chi$ of the form
\be \label{eq:Lagrangian}
{\cal L} =\sqrt{-g}\left[ \frac{R}{2} -  \frac{1}{2} \p_\m \phi \p^\m \phi - \frac{1}{2} e^{- b \phi} \p_\m \chi \p^\m \chi + V_o e^{-c \phi}  \right],
\ee
where we assume $b$ and $c$ to be constants, with the background equation of state given by $\epsilon = c^2/2.$ The background solution again lies along a constant $\chi$ curve. Because of the non-minimal coupling, the mode functions have to be defined slightly differently, the appropriate choice being  $v_s = a e^{- \frac{b}{2} \phi}  \delta \chi,$ whose linearised equation of motion in Fourier space is given by \cite{Fertig:2013kwa}
\be \label{eq:deltaseomtau_2}
v_s''  + \left[ k^2 - \frac{a''}{a}  - \frac{b^2}{4}\phi'^2  - \frac{b}{2} a^2  V_{,\phi}  \right] v_s =0\,,
\ee
where $k$ denotes the comoving wavenumber of the fluctuation mode. Up to an irrelevant global phase the solution is again a Hankel function
\be
v_s = \frac{\sqrt{-\pi\tau}}{2} H_\nu^{(1)}(-k\tau) \qquad \text{with} \quad \nu = \frac{3}{2} + \frac{(b-c)c}{c^2-2}\,.
\ee
At late times and on large scales, the entropy perturbations then scale as
\be \label{eq:modefunctionlate}
v_s \propto k^{-\nu} (-\tau)^{1/2-\nu}  \qquad (|k\tau| \ll 1).
\ee
If we define the fractional difference between $b$ and $c$ to be given by $\Delta \equiv \frac{b}{c}-1,$ then the spectral index can be written as
\be
n_s = 4 -  2 \nu = 1 - 2 \Delta \frac{\epsilon}{(\epsilon -1)}\,. \label{eq:spectrum}
\ee
For $b=c$ one obtains an exactly scale-invariant spectrum, while if $b$ is larger than $c$ by about one to two percent, we obtain $n_s = 0.97,$ in good agreement with Planck data \cite{Ade:2015xua}. Note that for the ekpyrotic phase to come to an end, $\ep$ has to decrease which naturally leads to a slight red shift in the spectrum for the most symmetric case where originally $b = c$ \cite{Fertig:2013kwa}, and this also generates a small negative running of the scalar spectral index \cite{Lehners:2015mra}.

An interesting feature of these models is that around the $\phi$-driven background we consider, they do not contain any terms cubic in $\delta \chi$ in the expansion of the Lagrangian \eqref{eq:Lagrangian}. Thus the second order correction vanishes, and the intrinsic non-Gaussianity is precisely zero \cite{Fertig:2013kwa}. Note that these models can be generalised to a large class of models with essentially identical properties, where the non-minimal coupling function $e^{b\phi}$ is generalised to a largely arbitrary function $\Omega^2(\phi)$ \cite{Ijjas:2014fja}. However, we will assume that the coupling function reverts to canonical towards the end of the ekpyrotic phase, so that during the bounce phase $\chi$ has a canonical kinetic term. As argued in \cite{Fertig:2015ola} this is both natural and necessary in order to make the conversion process tractable. 

\subsection{Conversion After the Bounce} \label{sec:ek_afterbounce}

So far we have described the kind of perturbations that are created by different kinds of ekpyrotic phases. These perturbations will pass though the bounce as described in Sec.~\ref{section:bounce}. At some point, the entropy perturbations need to be converted into curvature perturbations. We will now describe the evolution of the perturbations from the end of ekpyrosis to the end of the conversion process, assuming that the bounce occurs in between these phases. 

As we have just seen, at the end of the ekpyrotic phase the curvature perturbations are negligible, while the entropy perturbations take the form
\bea
{\delta s}^{(1)} = \delta s_{ek} \, &,& \qquad 
\dot {\delta s}{}^{(1)}= -\frac{\delta s_{ek}}{t_{ek}}\\
{\delta s}^{(2)}= \frac{\kappa_3 \sqrt{\ep} }{8} \de s_{ek}^2 \, &,& \qquad 
\dot {\delta s}{}^{(2)}=-\frac{\kappa_3 \sqrt{\ep}}{4} \frac{\de s_{ek}^2}{t_{ek}}
\eea
For our numerical evaluations we choose $\delta s_{ek}=10^{-5}.$ Note that the non-minimally coupled models simply correspond to specifying $\kappa_3=0.$

Before turning to specific examples, let us present the relevant equations describing the conversion process. The curvature perturbation on large scales and in comoving gauge can be expressed very simply as  \cite{Lyth:2004gb,Buchbinder:2007at,Lehners:2009qu}
\be
\dot{\cal{R}}=\frac{2H \de V}{\dot \sigma^2-2 \de V}
\ee
where $\de V=V(t,x^i)-\bar{V}(t)$. Expanding up to second order yields the equation that we need, namely
\bea \label{zetadot}
\dot{\cal{R}}&=&-\frac{2H }{\dot \sigma}  \dot \theta \, \de s+\frac{H }{\dot \sigma^2} \left(V_{ss}+4 \dot \theta^2  \right)(\de s)^2-
\frac{V_\sigma}{\dot \sigma } \de s \dot{\de s}.
\eea
It describes how the entropy perturbations act as a source for the large-scale curvature modes. Incidentally, note that this equation implies that already during the ekpyrotic phase there can be a contribution at second order to the curvature perturbation, even if no curvature perturbation is sourced at linear order. However, it turns out that this contribution is small (and actually exactly zero for the non-minimally coupled models), and hence we can ignore it here. In order to characterise the level of non-Gaussianity that is eventually obtained, it is customary to introduce the (local) non-Gaussianity parameter $f_{NL},$ defined via the relation
\be
{\cal{R}}={\cal{R}}^{(1)}+\frac{3}{5}f_{NL} \left( {\cal{R}}^{(1)} \right)^2\,.
\ee
This parameter specifies the 3-point function of the curvature perturbations in the so-called squeezed limit. Physically speaking, it characterises the extent to which a long-wavelength perturbation causes a modulation in the power spectrum of shorter-wavelength fluctuations. The current observational limit on $f_{NL}$ obtained by the Planck collaboration is \cite{Ade:2015ava}
\be
f_{NL} = 0.8 \pm 10.0 \,,
\ee
at the $95\%$ confidence level. It will be useful to keep this currently allowed range in mind when discussing specific models below.

In order to model the conversion, we will add a term to the potential, which will induce a bending of the background trajectory%
\footnote{It is also possible for an unstable ekpyrotic potential by itself to induce a bending of the trajectory and cause a subsequent conversion of perturbations \cite{Koyama:2007mg, Koyama:2007ag, Buchbinder:2007ad}. However, it turns out that such models lead to very large non-Gaussianities \cite{Buchbinder:2007at, Koyama:2007if, Lehners:2009ja}, in disagreement with observations \cite{Ade:2015ava}.}.
In a more fundamental context such a term is in principle calculable (see e.g.~the heterotic M-theory embedding of the cyclic universe \cite{Lehners:2007nb}). Here we will choose our total potential to be given by
\be \label{eq:unstpot}
V(\f)=-\frac{2V_o}{e^{-\sqrt{2\ep} \f }+e^{\sqrt{2 \ep} \f }} \left( 1+ \frac{1}{2} \ep \c^2+\frac{1}{3!}\ep^{3/2} \kappa_3 \chi^3  \right)+V_{rep}
\ee
with
\be
V_{rep}= 10^{-4} \, V_o \times e^{-5 \left[ \sin(\pi/10)\f-\cos(\pi/10) \c - 2\right]^2}.
\ee
Note that the repulsive potential is simply a smooth (Gaussian-shaped) barrier oriented at an angle to the background trajectory. The bounce occurs at $\f=0$ and the conversion process begins at around $\f=-2$. We should comment on the overall scale that we chose for the repulsive part of the potential: as has been discussed in previous works \cite{Lehners:2007wc,Lehners:2008my}, the most important aspect of the conversion process is the overall efficiency of the conversion, i.e.~how much of the initial entropy perturbations is converted into curvature perturbations. Inefficient conversions lead to little structure formation and a very wide range of predictions for non-Gaussian corrections. By contrast, the predictions narrow to a much smaller range for efficient conversions, and moreover they lead to greater structure formation. The amplitude of the entropy perturbation generated via ekpyrosis depends crucially on the energy scale reached by the potential. This energy scale must be rather large (i.e.~not too far below the Planck scale) in order for the perturbations to obtain an amplitude in agreement with observations. Hence the conversion process must indeed be efficient, as one cannot assume that the potential reaches even larger energy scales (at which point the effective description used here would certainly break down). For these reasons we will focus on efficient conversions, and this translates into our choice for the overall scale of the repulsive potential. An important open problem is to justify this assumption in a more fundamental context.

\begin{figure}[htbp]
\begin{minipage}{0.5\textwidth}
\includegraphics[width=0.92\textwidth]{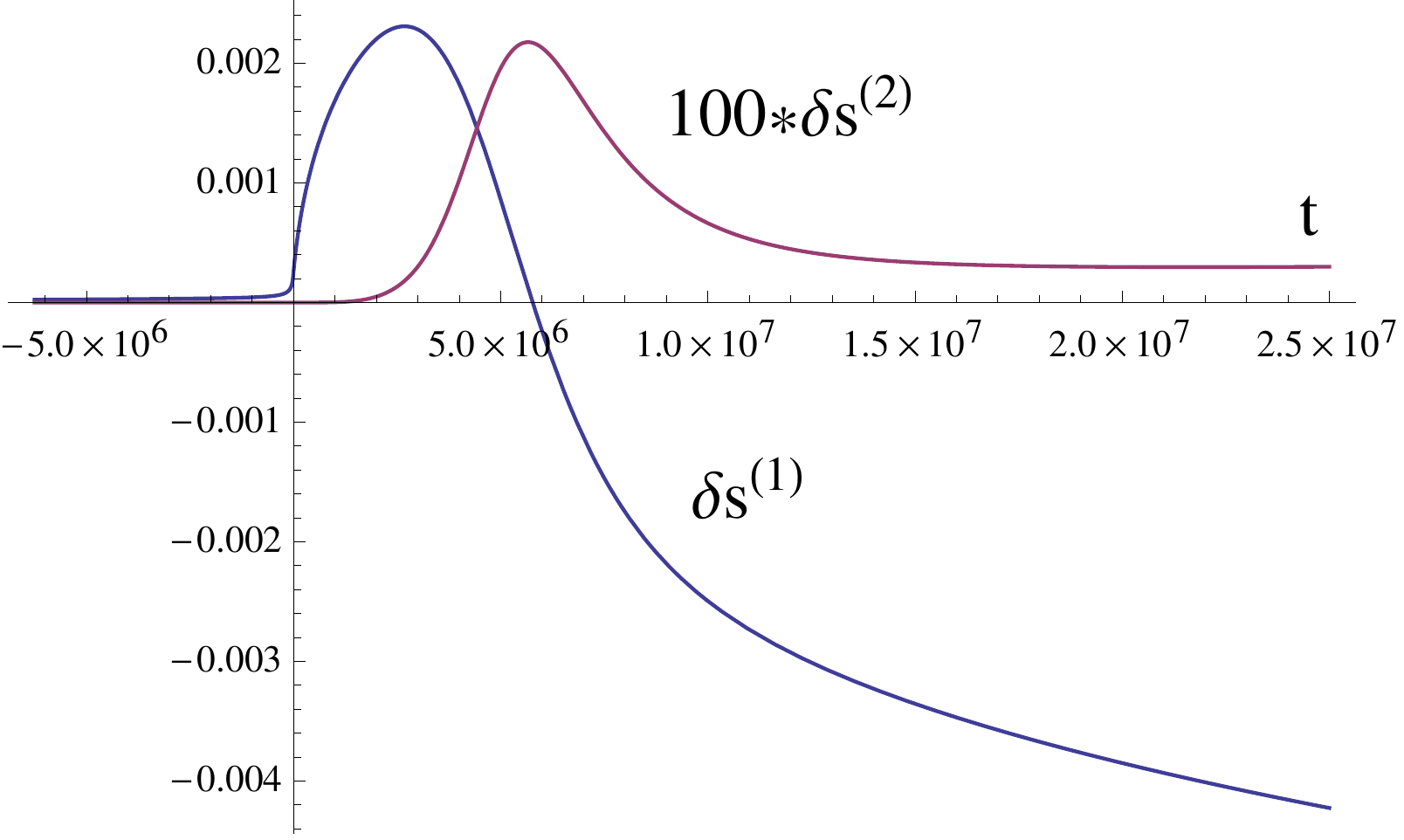}
\end{minipage}%
\begin{minipage}{0.5\textwidth}
\includegraphics[width=0.92\textwidth]{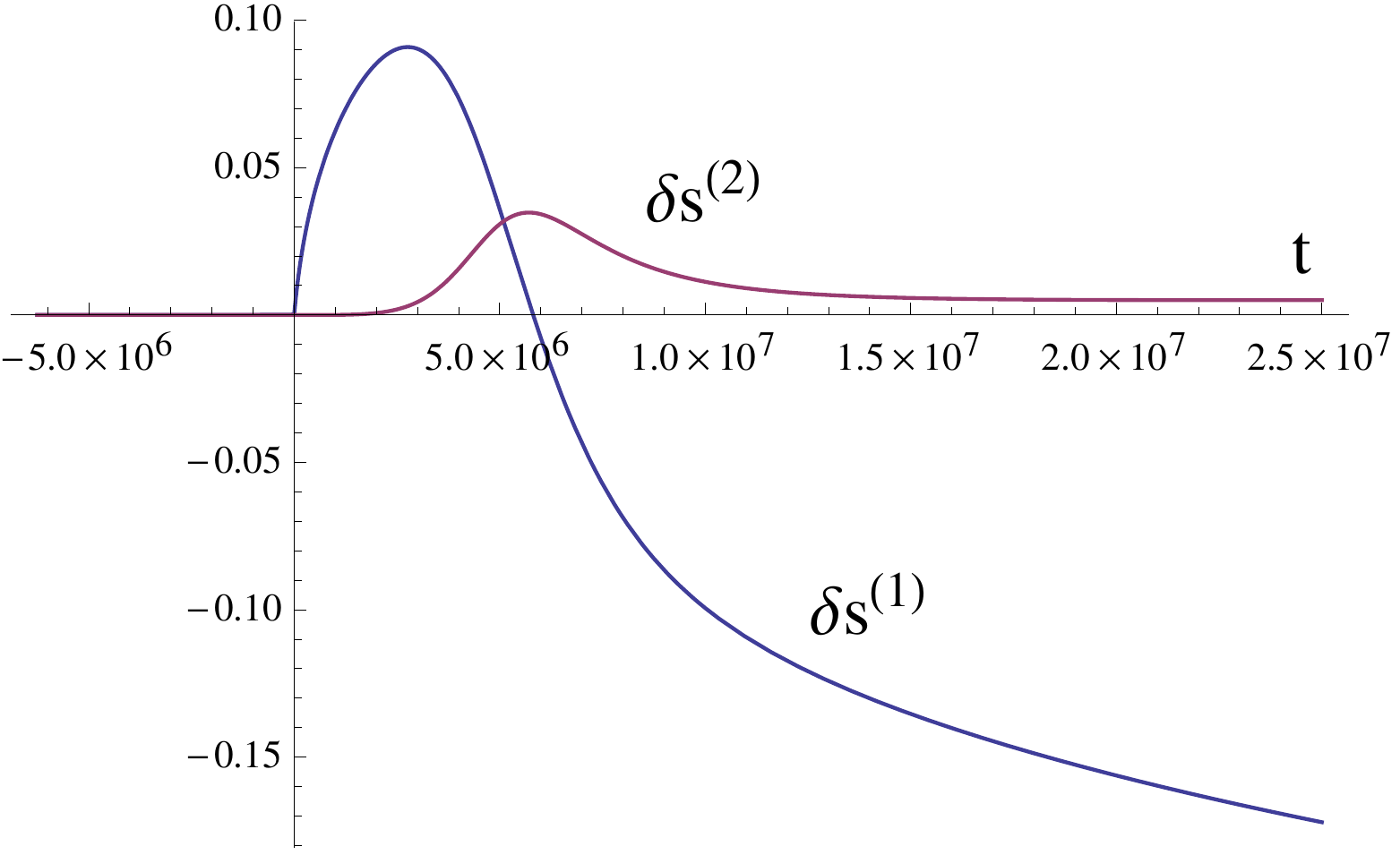}
\end{minipage}%
\caption{\footnotesize \hangindent=10pt
The evolution of the linear and second order entropy perturbations from after the bounce through the conversion phase. Here we have taken $\kappa_3=0.$ The left panel corresponds to having a stable potential \eqref{eq:stpot} during the bounce (i.e.~a flat transverse direction), while the right panel shows the case of the unstable potential \eqref{eq:unstpot}.} 
\label{fig:s1s2}
\end{figure}

\begin{figure}[htbp]
	\begin{minipage}{0.5\textwidth}
		\includegraphics[width=0.92\textwidth]{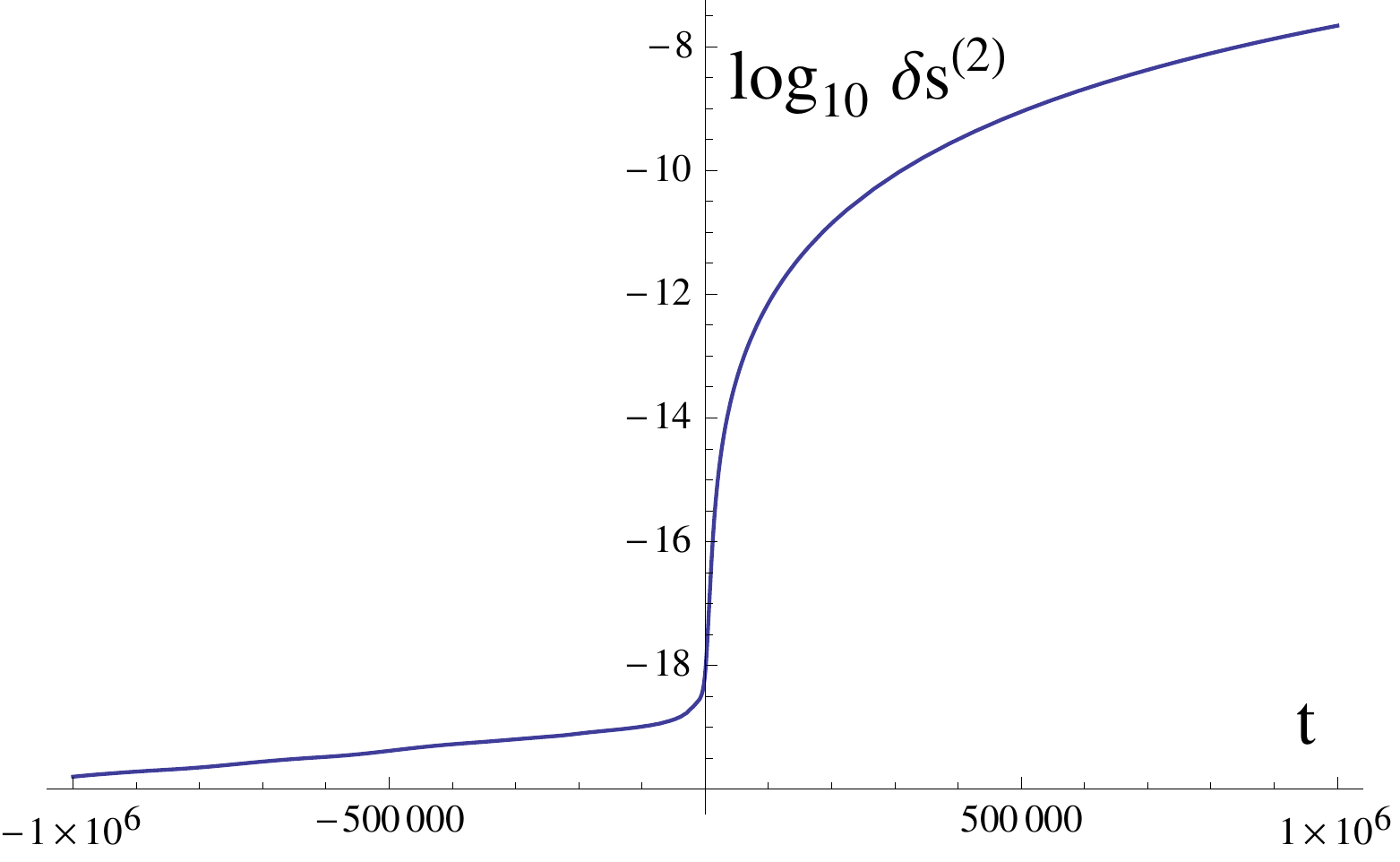}
	\end{minipage}%
	\begin{minipage}{0.5\textwidth}
		\includegraphics[width=0.92\textwidth]{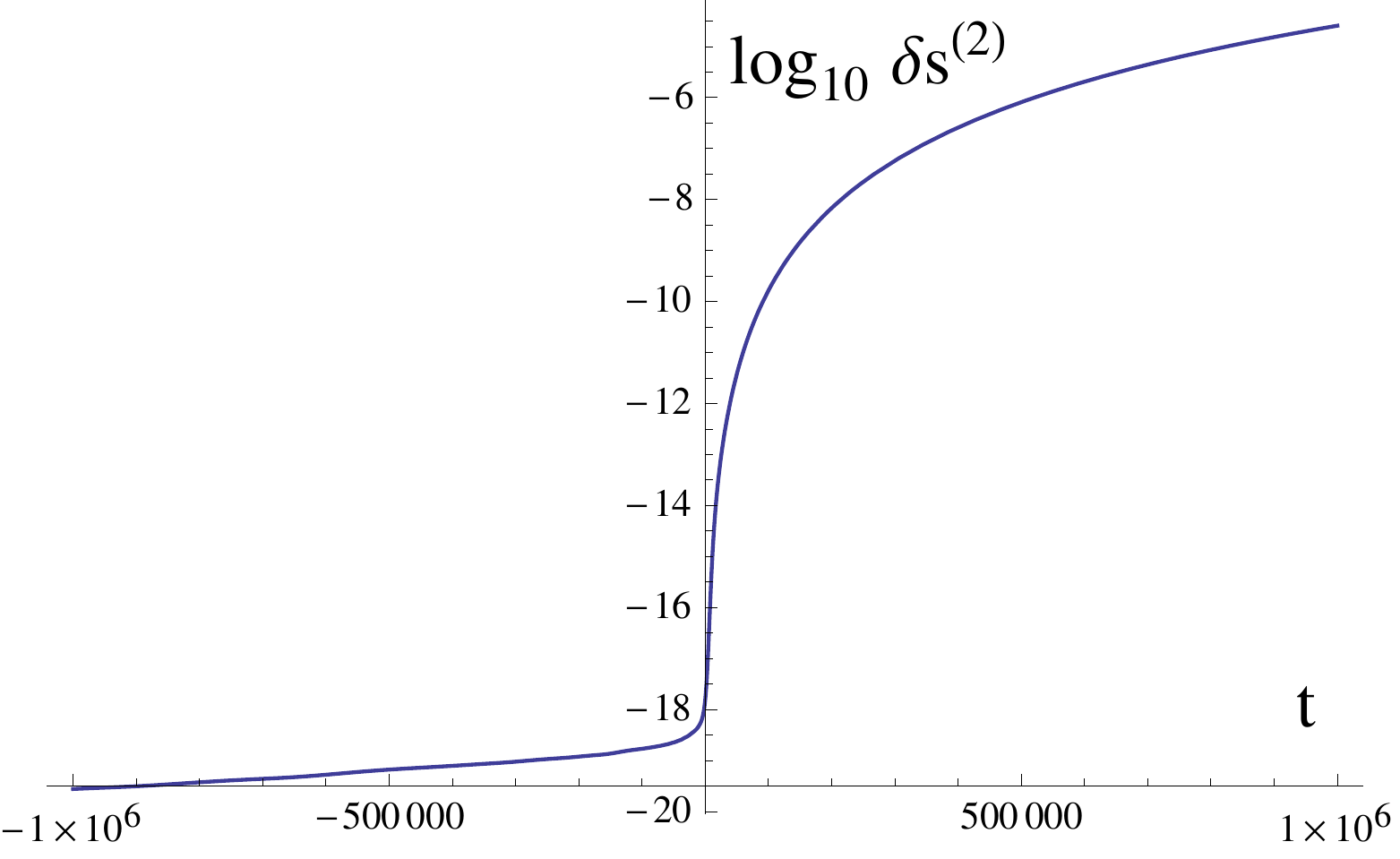}
	\end{minipage}%
	\caption{\footnotesize \hangindent=10pt
The left panel shows the logarithm of the second order entropy perturbation for a bounce with the stable potential \eqref{eq:stpot} and a preceding kinetic phase. The right panel shows the analogous plot for a bounce with the unstable potential \eqref{eq:unstpot}. The logarithm shows more clearly the overall amplification through the kinetic and bounce phases.}
	\label{fig:s2kin}
\end{figure}

We have performed a series of numerical simulations, following the entropy and curvature perturbations through the kinetic and bounce phases until after the conversion period. We will simply show a few typical examples. Figure~\ref{fig:s1s2} shows the evolution of the linear and second order entropy perturbations from after the bounce through the conversion phase. The left panel corresponds  to having a stable potential during the bounce (i.e.~a flat transverse direction)
\be \label{eq:stpot}
V(\f)=-\frac{2V_o}{e^{-\sqrt{2\ep} \f }+e^{\sqrt{2 \ep} \f }} +V_{rep},
\ee
while the right panel shows the case of the unstable potential \eqref{eq:unstpot}, where we have used the background parameters given by \eqref{eq:parameters}, namely $\epsilon = 10, V_o=2\times 10^{-8}.$ The linear entropy perturbation grows significantly during the bounce phase, and starts oscillating during the conversion. The second order perturbation grows as a consequence of the growth of the linear perturbation, since the linear perturbation acts as its source. As a result, the growth of the second order perturbation lags a little behind the growth of the linear perturbation, but the overall growth is very significant, as shown in more detail in Fig.~\ref{fig:s2kin}.

\begin{figure}[tbp]
\begin{minipage}{0.5\textwidth}
\includegraphics[width=0.92\textwidth]{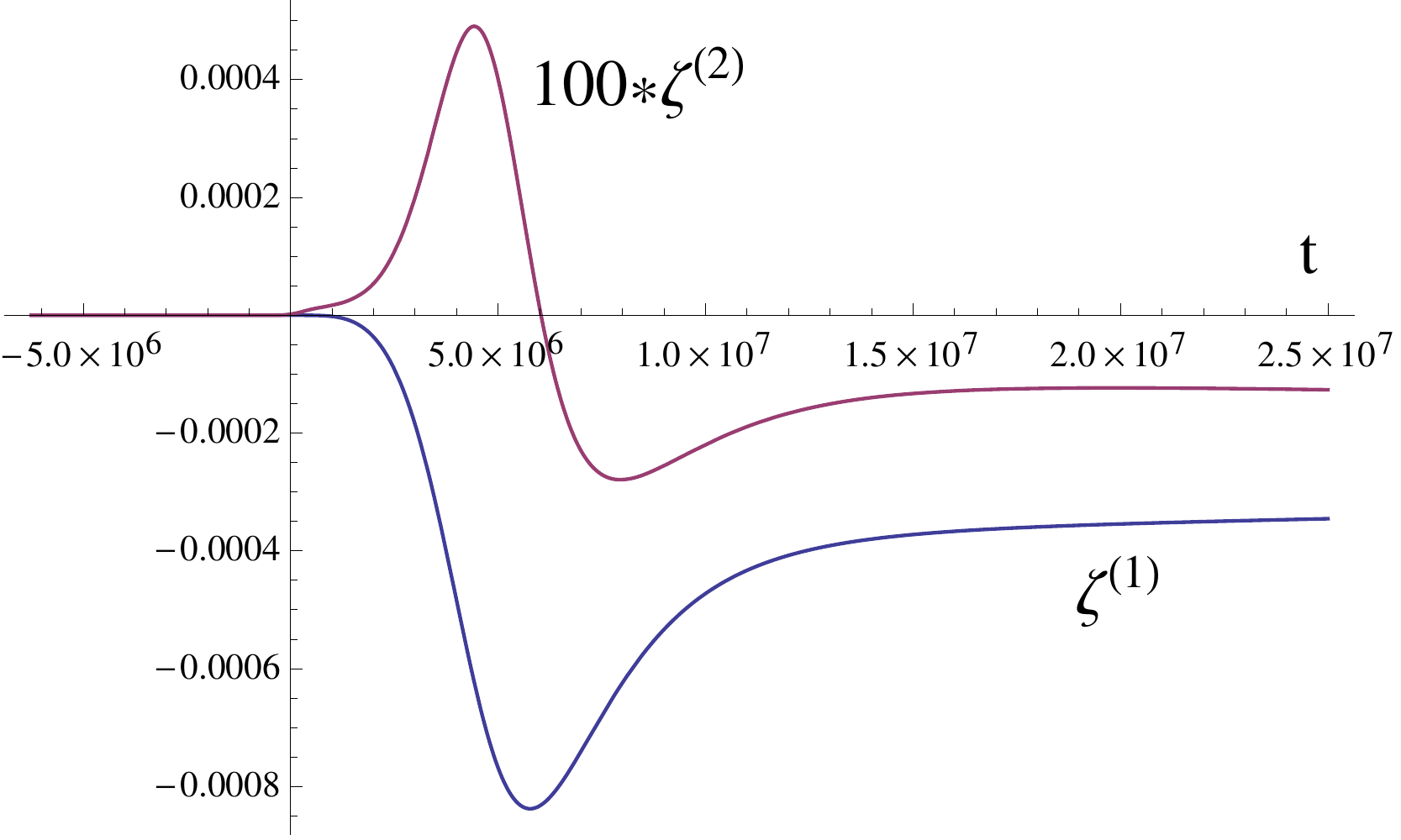}
\end{minipage}%
\begin{minipage}{0.5\textwidth}
\includegraphics[width=0.92\textwidth]{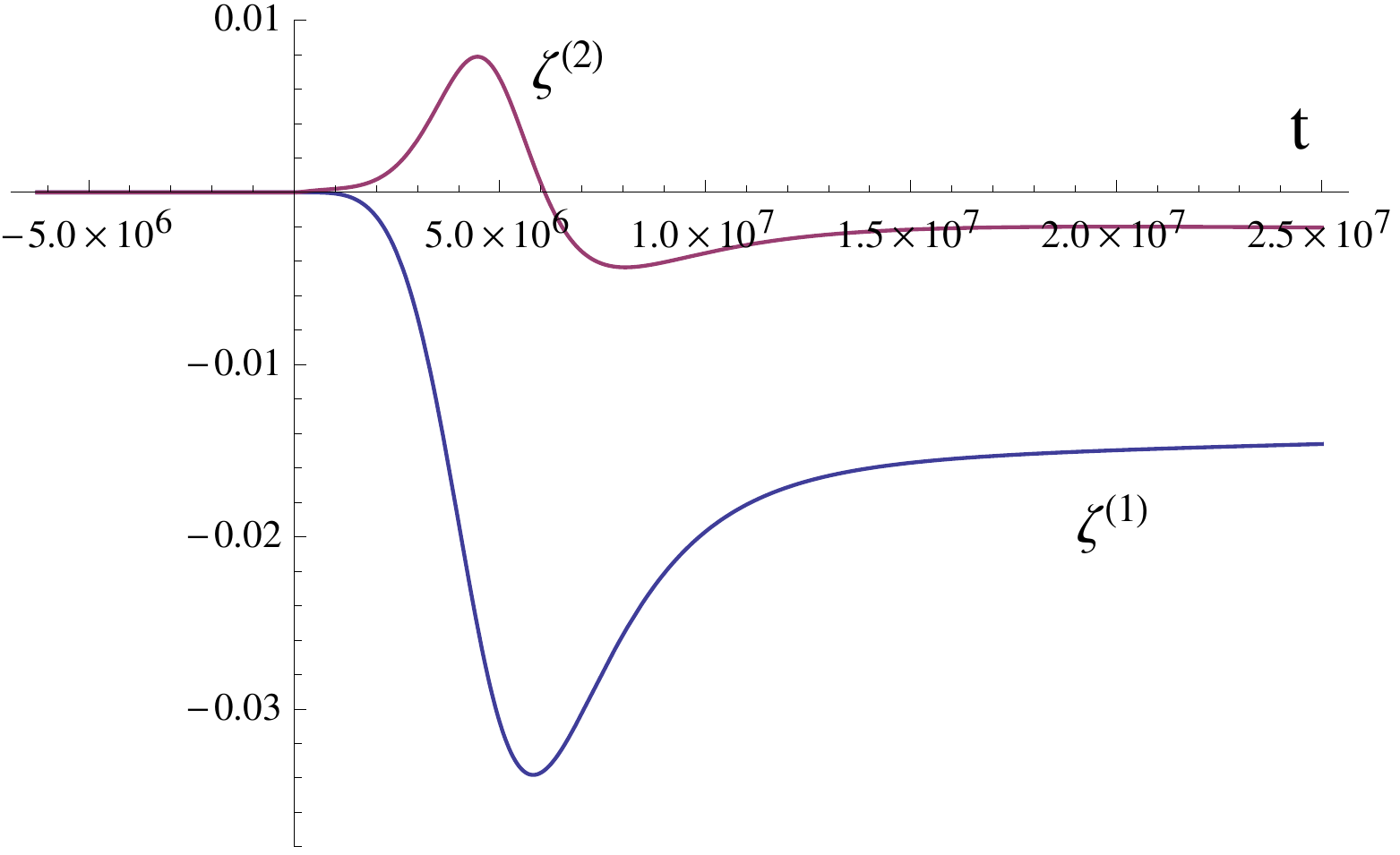}
\end{minipage}%
\caption{\footnotesize \hangindent=10pt
The evolution of the linear and second order curvature perturbations from after the end of ekpyrosis through the conversion phase. Here we have again taken $\kappa_3=0.$ The left panel corresponds  to having a stable potential \eqref{eq:stpot} during the bounce (i.e.~a flat transverse direction), while the right panel shows the case of the unstable potential \eqref{eq:unstpot}.} 
\label{fig:z1z2}
\end{figure}

The curvature perturbations at linear and second order are plotted in Fig.~\ref{fig:z1z2}, where one clearly sees the period of the conversion. At late times, the field space trajectory bends less and less, and the curvature perturbations approach constant values, as expected. 

For the specific example shown in the left panels of Figs.~\ref{fig:s1s2}--\ref{fig:z1z2}, the potential during the bounce is stable. In this example, the final value of the curvature perturbation ${\cal R},$ its ratio to the entropy perturbation at the beginning of the conversion phase (i.e.~the time at which the trajectory starts bending significantly) and the value of the final non-Gaussianity parameter $f_{NL}$ are given by 
\bea
{\cal R}_{final} &=&-3.5 \times 10^{-4}\,,\\
\frac{{\cal R}_{final}}{\delta s_{conv-beg}} &=& -0.18\,,\\
f_{NL} &=&-4.0\,.
\eea
For the time interval from $t_1=1.45 \times 10^{6}$ to $t_2=10.5 \times 10^{6}$ the background evolves by one e-fold, i.e.~$aH$ changes by a factor of $e,$ and during this time $95 \%$ of the total conversion takes place. Hence this is clearly a very efficient, yet smooth, conversion. The analogous example for the case with an unstable potential during the bounce phase is shown in the right panels of Figs.~\ref{fig:s1s2}--\ref{fig:z1z2}. Here we obtain the values
\bea
{\cal R}_{final} &=&-1.5 \times 10^{-2}\,,\label{example2}\\
\frac{{\cal R}_{final}}{\delta s_{conv-beg}} &=& -0.20\,,\\
f_{NL}&=&-3.6\,.
\eea
The conversion period is the same as above, and this is an equally efficient conversion. As expected, the final curvature perturbation is however much larger, by almost two orders of magnitude in the present case. Nevertheless, the value of the non-Gaussianity parameter $f_{NL}$ remains of the same order, as the second order perturbation has been amplified correspondingly. 

\begin{figure} 
	\centering
	\includegraphics[width=0.5\textwidth]{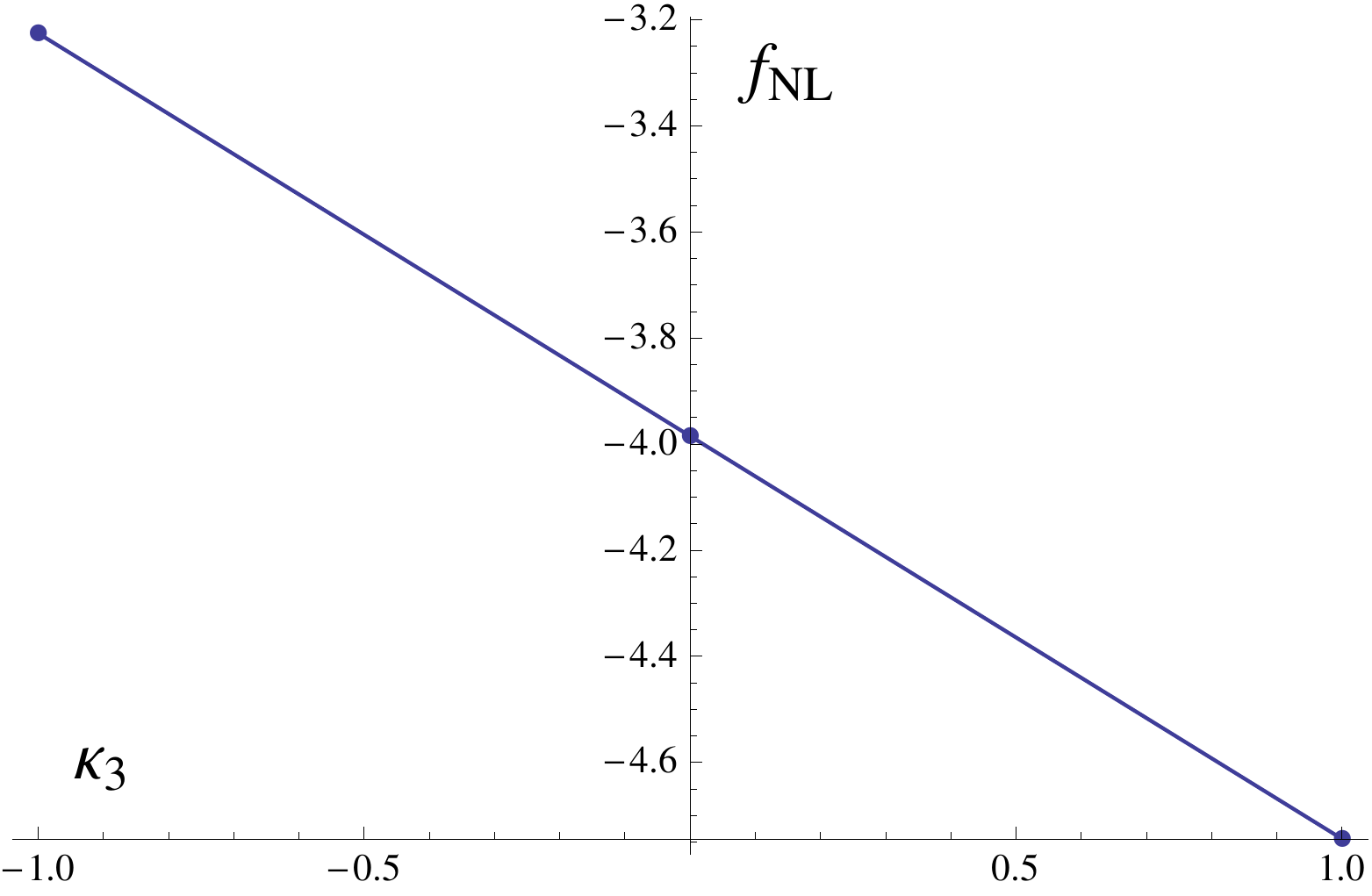}
	\caption{\footnotesize \hangindent=10pt
The dependence of $f_{NL}$ on the intrinsic non-Gaussianity (parameterised by $\kappa_3 $) is seen to be surprisingly small when converting after the bounce. The example shown here has a stable potential during the bounce.}
	\protect
	\label{fig:no_unstable_kappa}
\end{figure}

In the examples discussed so far, the entropy perturbation was perfectly Gaussian before the conversion process. This is the relevant case for the non-minimally coupled ekpyrotic models. However, in ekpyrotic models with an unstable potential, there can already be a significant intrinsic non-Gaussianity in the entropy perturbation, parameterised by $\kappa_3$ and depending on the tilt of the potential \eqref{eq:unstable3}. The dependence of the final value of $f_{NL}$ on $\kappa_3$ is shown in Fig.~\ref{fig:no_unstable_kappa}, for the example where the bounce is stable. Interestingly, the dependence on $\kappa_3$ is very weak, as the change in $f_{NL}$ is smaller than $2$ for $-1< \kappa_3 < +1.$ This may be understood as follows: as the entropy perturbations grow during the bounce phase, the second order part is sourced more and more strongly by the terms quadratic in the linear perturbation. Thus, even though the intrinsic second order term is also amplified (roughly like the linear term), its relative importance compared to the square of the linear term is lessened. Thus, for these models where the conversion takes place after the bounce, the main contribution to the non-Gaussianity comes from the non-linearity of the conversion process itself, and not from a possible intrinsic non-Gaussianity generated during the contraction phase. This is precisely the opposite situation to that described in \cite{Lehners:2009qu}, where the conversion process was analysed for large intrinsic non-Gaussianity and neglecting the non-linearities of the conversion process. Thus, converting after the bounce rather than before brings with it a significant change in the implications of the conversion process.

\subsection{Conversion Before the Bounce}

\begin{figure} 
	\centering
	\includegraphics[width=0.5\textwidth]{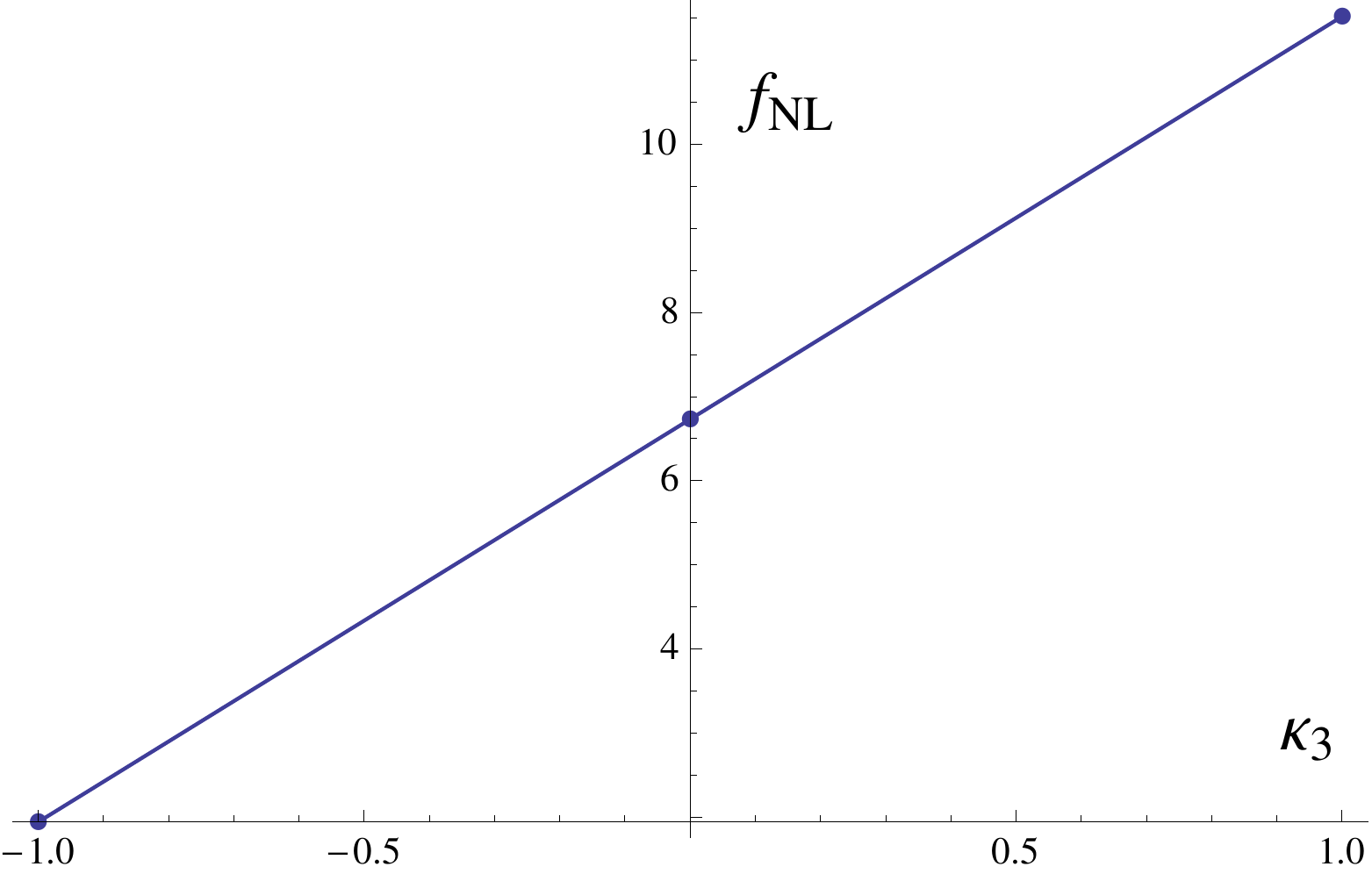}
	\caption{\footnotesize \hangindent=10pt
The dependence of $f_{NL}$ on the intrinsic non-Gaussianity (parameterised by $\kappa_3 $) for a model of conversion before the bounce. Note that the slope has the opposite sign compared to the case when converting after the bounce: this is due to the fact that here the universe contracts whereas it expands in the other case.}
	\protect
	\label{fig:no_bounce_kappa}
\end{figure}

It may be useful to contrast the new results that we just described with the old case of having the conversion occur during the phase of kinetic contraction before the bounce. This conversion process has been discussed at length in \cite{Lehners:2007ac,Lehners:2009qu,Lehners:2010fy}, hence we do not need to go into details here. Following \cite{Fertig:2015ola}, we take the repulsive potential to be
\be
V_{rep}= \frac{12 \times 10^{-9}}{\left[ \sin(\pi/6)\f + \cos(\pi/6) \c +2\right]^2}\,,
\ee
and the initial conditions for the numerical evolution are given by
\bea
t_o=-1000 \quad&,&\quad a_o =1 \\
\phi_o=-\sqrt{\frac{2}{3}}-4.5 \quad&,&\quad
\dot \phi_o=\sqrt{\frac{2}{3}} \, \frac{1}{|t_o|}\\
\delta s_o=10^{-5} \quad&,&\quad
\dot{\delta s_o}=\frac{10^{-5}}{|t_o|}\,.
\eea
Then we find
\bea
{\cal R}_{final} &=& 2.3 \times 10^{-6}\,,\\
\frac{{\cal R}_{final}}{\delta s_{conv-beg}} &=& 0.14\,,\\
f_{NL} &=& 6.7\,.
\eea
For this example, $87 \%$ of the conversion take place over one e-fold of evolution from around $t_1=-440$ to $t_2=-70$; thus for this particular example the conversion is slightly less efficient than it was in the examples above, but nevertheless comparable. Due to this difference in efficiency, the value of $f_{NL}$ is slightly higher. The dependence of $f_{NL}$ on $\kappa_3$ is shown in Fig.~\ref{fig:no_bounce_kappa}. Note that this dependence, which is well parameterised by the phenomenological formula $f_{NL} \approx \frac{3}{2}\sqrt{\ep} \kappa_3+5$ \cite{Lehners:2010fy}, is vastly more significant in the present case than when converting after the bounce and therefore a wider range of values is obtained. 

\subsection{Implications for Models and Observations}

The main implication of having a conversion phase during the expansion phase after the bounce, rather than before the bounce, is that the final amplitude of the curvature perturbation is significantly enhanced. As we saw with the help of specific examples, for the case of a bounce with a stable potential the curvature perturbation is approximately two orders of magnitude larger than when the conversion takes place before the bounce. For an unstable potential, the growth is even more significant, by almost four orders of magnitude in the example shown around Eq.~\eqref{example2}. It is useful to recall from Eq.~\eqref{entropylineargenerated} that at the end of the ekpyrotic phase the (Fourier mode of the) entropy perturbation reaches a value 
\be
\delta s_k (t_{ek-end}) \approx \frac{|\ep V_{ek-end}|^{1/2}}{\sqrt{2}k^{\nu}}\,.
\ee
For the case where the conversion takes place during the contraction phase before the bounce, this leads to the following root mean square value for the curvature perturbation (where we have used the approximation ${\cal R} \approx \delta s_{ek-end}$ appropriate for such models, see \cite{Lehners:2009qu}),
\be
\langle{\cal R}^2 \rangle \approx  \int \frac{\d^3k}{(2\pi)^3}\frac{\ep V_{ek-end}}{50 k^{2\nu}}=\int \frac{\d k}{k}\frac{\ep V_{ek-end}}{100 \pi^2 }k^{n_s-1}\,.
\ee
Matching to the observed value of $\Delta^2_{{\cal R}} = 2.4 \times 10^{-9}$ \cite{Ade:2015xua} leads to an estimated depth of the potential of $| V_{ek-end}| \approx (10^{-2}M_{Pl})^4,$ around the grand unified scale. For conversion after the bounce, we obtain an enhancement of the curvature perturbation by approximately two to four orders of magnitude. This would mean that the potential does not have to become quite so deep, and the energy scale that the potential must reach is around $10^{-4}$ or $10^{-3}$ in reduced Planck units. 

Despite the fact that the amplitude of the curvature perturbation is significantly enhanced by the conversion process, the level of non-Gaussianity remains similar (though as we have seen the value of the local non-Gaussianity parameter $f_{NL}$ is now mostly determined by the conversion process and not by the intrinsic level of non-Gaussianity produced during the ekpyrotic phase). Our numerical examples suggest that for efficient conversion processes typical values lie in the range
\be
-5 \lesssim f_{NL} \lesssim +5\,,
\ee
while a little less efficient conversions lead to values of ${\cal O}(10).$ This is a very interesting range of values, consistent with current bounds on $f_{NL},$ which are $f_{NL} = 0.8 \pm 10.0$ at the $2\sigma$ level \cite{Ade:2015xua}, but within the reach of not-too-distant future experiments. In particular, the Square Kilomater Array is expecting to reduce the error on local non-Gaussianities to less than $1$ \cite{Gaensler}, at which point one would expect to measure the presently discussed values.

Some of the features we have assumed in this work, in particular having two scalar fields and having a bend in the field space trajectory, arise automatically in the heterotic M-theory embedding of the cyclic model of the universe \cite{Lehners:2006pu,Lehners:2006ir,Lehners:2007ac}. The big difference is that the big bang is not replaced by a non-singular bounce, but rather by a collision of branes \cite{Khoury:2001wf} (which remains ill-understood to date; for the sake of the present discussion we will assume that it can effectively be replaced by a bounce). In the model inspired by heterotic M-theory, the conversion of entropic into adiabatic modes occurs because the negative-tension brane gets repelled by a time-like singularity which in moduli space translates into a bending of the trajectory. As argued above, the associated conversion needs to be smooth in order to be efficient, and in order for non-Gaussianities to remain small. However, there is a potential difficulty here: the ekpyrotic phase dilutes all matter, so that one would expect that after the ekpyrotic phase the conversion ought to be very sharp. In that case the amplitude of the perturbations would be small and $f_{NL}$ huge. In the present work we see a possible resolution, namely that at the brane collision the universe reheats and new matter is produced. Moreover, after the brane collision the trajectory in moduli space is just the reverse of the trajectory before the collision. Hence, the dominant part of the conversion is more likely to happen after the bang, and this would naturally favour smaller non-Gaussianity. 

Another consequence is that at the brane collision matter is produced on both branes and one might imagine that there is a probability distribution (ultimately due to the quantum nature of particle creation) for how much matter ends up on the positive tension brane and how much on the negative tension brane. If more matter is produced on the negative tension brane, then the conversion will be smoother (since it is precisely the dynamics of the negative-tension brane that causes the conversion), implying more structure and less non-Gaussianity at the same time. Moreover, matter on the negative tension brane would be seen as dark matter. Hence, in such a model one might obtain a correlation between small non-Gaussianity and significant amounts of dark matter, which would provide a new link between two cosmological observations. This is certainly a very attractive prospect, although to explore it further requires a better understanding of brane collisions.

\section{Matter Bounce Model}
\label{section:matter}

Scale-invariant perturbations are also generated from vacuum quantum fluctuations in a contracting universe dominated by a pressureless fluid.  In this section, we will study a two-field realization of this scenario, with a massive scalar field $\chi$ and an ekpyrotic scalar field $\phi$, as first considered in \cite{Cai:2013kja}.  The bounce is again caused by the ekpyrotic scalar field $\phi$ becoming a ghost condensate at high energy densities, as described above in Sec.~\ref{section:bounce}.

Therefore, the total gravitational and matter action is again of the form \eqref{eq:actiongeneral}, with
\be \label{mb-pot}
V(\phi, \chi) = \frac{1}{2} m^2 \chi^2 - V_o \, e^{c \phi},
\ee
and allowing for possible modifications of the form of the potential energy near the bounce point, again as described in Sec.~\ref{section:bounce}.

Away from the bounce where the ghost condensate contributions are negligible, the background equations of motions are the Friedmann and Raychaudhuri equations,
\begin{align}
3H^2 &= \frac{1}{2} \dot \phi^2 +\frac{1}{2} \dot \chi^2 + V(\phi, \chi), \\
\dot H &= -\frac{1}{2} \dot \phi^2 - \frac{1}{2} \dot \chi^2,
\end{align}
together with the two Klein-Gordon equations,
\begin{align}
\ddot \phi + 3H \dot \phi - c V_o e^{c \phi} &= 0, \\
\ddot \chi + 3H \dot \chi + m^2 \chi &= 0, \label{kg-chi}
\end{align}
one for each scalar field.

Following \cite{Cai:2013kja}, we will set initial conditions in the contracting branch of the cosmology well before the bounce, such that the massive field $\chi$ is oscillating in its potential, while the ekpyrotic field $\phi$ starts at rest (or nearly) at a very large negative field value so that its contribution to the potential energy is initially negligible.

In this way, at first the massive field dominates the dynamics and acts, on average, as a pressureless matter fluid.  During this time, as we shall see more explicitly below, the entropy perturbations that exit the horizon will become scale-invariant and they shall in turn generate scale-invariant curvature perturbations.  After some time, the ekpyrotic field will near $\phi = 0$ where its contribution to the potential energy becomes important, at which point the space-time will undergo ekpyrotic contraction, hence avoiding the anisotropy instability.  Finally, a bounce will occur when the ghost condensate corrections to the action become important. While the curvature perturbations remain frozen during the ekpyrotic and bounce phases, the entropy perturbations can grow and, as we shall see, if the entropy perturbations are efficiently converted to curvature perturbations after the bounce, the entropy perturbations will provide the dominant contribution to the curvature perturbations.  The details of the background dynamics for this cosmological scenario are given in Sec.~\ref{ss.mb-back}, while the calculation of the evolution of the curvature and entropy perturbations in this background space-time is presented in Secs.~\ref{ss.mb-pert-dust} and \ref{ss.mb-pert-ekp}. The results of converting the entropy perturbations to curvature perturbations after the bounce are presented in Sec.~\ref{ss.conv-mb}.   For the sake of completeness we will recall a number of relevant results from \cite{Cai:2013kja} in Secs.~\ref{ss.mb-back}, \ref{ss.mb-pert-dust} and \ref{ss.mb-pert-ekp}.

\subsection{Background Evolution}
\label{ss.mb-back}

In the first phase of contraction, the massive scalar field is oscillating while the ekpyrotic scalar field is slowly moving in a region where its contribution to the potential is essentially vanishing and so the contribution of the ekpyrotic field to the total energy density and pressure is negligible.  The background evolution is then approximately that of a pressureless matter field.  Since the matter-dominated phase will be followed by an ekpyrotic phase with the transition occurring at the ``equality time'' $t_e$, it is convenient to express the scale factor as
\be \label{eq:a_matter}
a(t) = a_o (t - t_o)^{2/3},
\ee
with $a_o = a_e (3 H_e / 2)^{2/3}$ and $t_o = t_e - 2 / 3 H_e$, where $a_e$ and $H_e$ are respectively the values of the scale factor and the Hubble rate at $t_e$.  The two scalar fields evolve as
\be
\chi(t) = \sqrt \fr{8}{3} \cdot \fr{\sin m (t-t_o)}{m (t-t_o)},
\ee
\be \label{phi-t}
\phi(t) = \fr{\sqrt{32}}{\sqrt{27} \, H_e (t-t_o)} + \phi_o,
\ee
with $\phi_o < 0$.  This solution holds so long as (i) $\phi(t) \ll 1$ so that the ekpyrotic potential is negligible, and (ii) $m^2 (t-t_o)^2 \gg 1$ so that the potential term in \eqref{kg-chi} dominates over the anti-friction term.  Here we choose the initial conditions so that the condition (i) fails before condition (ii), this means that the pressureless phase of contraction transitions directly to the ekpyrotic phase without any intermediate kinetic-dominated phase (where the effective equation of state is $\omega = 1$).

Note that at the transition time $t_e$ between the matter and ekpyrotic phases $\rho_\phi = \rho_\chi$, which implies that
\be \label{eq-vel}
\langle \dot\chi^2 \rangle = \fr{1}{2} \dot\phi^2,
\ee
where $\langle \cdot \rangle$ denotes the time average.  This relation determines the numerical prefactor to the time-dependent term in \eqref{phi-t}.

Once the ekpyrotic field dominates the dynamics the ekpyrotic attractor will be reached very rapidly, and at this point the scale factor and the ekpyrotic field evolve according to the scaling solutions \eqref{scalingsolution}, while the dynamics of $\chi$ is negligible compared to $\phi$ and it becomes a spectator field.  Finally, when the $\phi$ field becomes a ghost condensate a bounce occurs as explained in detail in Sec.~\ref{section:bounce}.

\subsection{Perturbations in the Matter-Dominated Phase}
\label{ss.mb-pert-dust}

A convenient set of gauge-invariant variables for perturbations in the scalar fields and the metric are the Mukhanov-Sasaki variables
\be
v_\chi = a \delta \chi + \fr{a \dot\chi}{H} \psi, \qquad
v_\phi = a \delta \phi + \fr{a \dot\phi}{H} \psi,
\ee
where $2 \psi$ is the perturbation in the metric component $g_{tt}$.  (For more details on these variables see e.g.~Appendix A of \cite{Cai:2013kja}.)  These variables are related to the co-moving curvature perturbation by
\be
\mR = \fr{H}{a} \cdot \fr{\dot\phi v_\phi + \dot\chi v_\chi}{\dot\phi^2 + \dot\chi^2},
\ee
and to the entropy perturbation by
\be \label{def-ds}
\delta s = \fr{1}{a} \cdot \fr{\dot\phi v_\chi - \dot\chi v_\phi}{\sqrt{\dot\phi^2 + \dot\chi^2}}.
\ee
The equations of motion for the Mukhanov-Sasaki variables in Fourier space are
\be \label{ms1}
v_\phi'' + \left( k^2 - a^2 c V_o e^{c \phi} - \fr{a''}{a} \right) v_\phi
- J_{\phi\phi} v_\phi - J_{\phi\chi} v_\chi = 0,
\ee
\be \label{ms2}
v_\chi'' + \left( k^2 - a^2 m^2 - \fr{a''}{a} \right) v_\chi
- J_{\chi\phi} v_\phi - J_{\chi\chi} v_\chi = 0,
\ee
where primes denote derivatives with respect to conformal time and
\be
J_{\alpha\beta} = \fr{1}{a} \fr{d}{dt} \left( \fr{a^3 \dot\alpha \dot\beta}{H} \right).
\ee

Approximate solutions to these equations can be obtained through the Born approximation \cite{Cai:2013kja}, where the $J$ source terms are treated as small perturbations.  During the pressureless phase of contraction, the contribution due to the ekpyrotic potential is negligible, $a^2 m^2 \gg a''/a$, and also $k^2 \ll a^2 m^2$ for the modes of interest for reasonable choices of $m$.

Assuming that the perturbations start in the vacuum quantum state, the zeroth order solution to the Mukhanov-Sasaki equations is
\be \label{born-zero}
v_\phi^{(0)} = \fr{1}{\sqrt{2 k}} \, e^{-i k (\eta - \eta_o)} \left( 1 - \fr{i}{k(\eta - \eta_o)}\right), \qquad
v_\chi^{(0)} = \fr{1}{\sqrt{2 am}} \, e^{-i am (\eta - \eta_o) / 3},
\ee
where the WKB approximation is used for $v_\chi$.  Note that due to the effect of the potential $m^2 \chi^2 / 2$, to zeroth order in the Born approximation the perturbations in the massive field $\chi$ do not feel the space-time curvature and hence do not become scale-invariant, nor are they significantly amplified.  On the other hand, the $v_\phi$ do feel the curvature once they exit the horizon, at which point they become scale-invariant.

Then, to first order in the Born approximation,
\be \label{born-exp}
v_\phi^{(1)} = (1 + A_\phi) v_\phi^{(0)} + B_\phi v_\chi^{(0)}, \qquad
v_\chi^{(1)} = (1 + A_\chi) v_\chi^{(0)} + B_\chi v_\phi^{(0)},
\ee
where the $A$ and $B$ depend on time and can be solved for from \eqref{ms1} and \eqref{ms2}.  Their exact form is not important here, the main point is that as a result of the source terms $J$, both $v_\phi$ and $v_\chi$ become scale-invariant when they exit their horizon.  (Since $v_\chi^{(0)}$ always has a very blue spectrum, it is not relevant observationally and therefore can be ignored.)

Importantly, in addition to being scale-invariant, the amplitude of the curvature and entropy perturbations are of the same order of magnitude at the time of equality $t_e$, i.e.~at the beginning of the ekpyrotic phase.  This can be seen from the fact that at $t_e$ we have, from \eqref{eq-vel}, $\dot\phi \sim \dot\chi$ and then the entropy perturbations \eqref{def-ds} can be rewritten as
\be
\delta s(t_e) \sim \fr{1}{a} \cdot \fr{\dot\chi v_\chi - \dot\phi v_\phi}{\sqrt{\dot\phi^2 + \dot\chi^2}}.
\ee
Generically, there will not be any exact cancellations between the two terms in the numerator, which will therefore be of the same order of magnitude as the sum between the two terms.  This gives exactly the curvature perturbation, up to a prefactor of order unity coming from the Friedmann equation $H^2 = (2 \langle \dot\chi^2 \rangle + \dot\phi^2) / 6$ (neglecting the ekpyrotic potential; at most it will change the prefactor but not its order of magnitude).  Therefore, the amplitudes of the curvature and entropy perturbations approximately match and their scale-dependence is the same,
\be \label{same-amp}
\delta s(t_e) \sim \mR(t_e).
\ee
Furthermore, a more careful calculation determing the precise prefactors in \eqref{born-exp} shows that the amplitude of both the curvature and the entropy perturbations in natural units is $H_e$, the relevant length scale at the equality time \cite{Cai:2013kja}.  Thus, in this cosmology at the beginning of the ekpyrotic phase of contraction,
\be
\mR_k(t_e) \sim \delta s_k(t_e) \sim H_e \, k^{-3/2}.
\ee

\subsection{Perturbations in the Ekpyrotic and Bounce Phases}
\label{ss.mb-pert-ekp}

Once the ekpyrotic phase begins, both the curvature and entropy perturbations of observational interest are well outside their horizon.  In this limit, the equation of motion for the curvature perturbations is simply
\be
\dot{\mR} = -\frac{2H }{\dot \sigma}  \dot \theta \, \de s,
\ee
while the equation of motion for the entropy perturbations to linear order is
\be \label{eom-ds}
\ddot{\de s} + 3 H \dot{\de s} + \left( V_{ss} + 3 \dot \theta^2  \right) \de s = 0.
\ee
During the ekpyrotic phase $\dot\phi \gg |\dot\chi|$ and as a consequence $\dot\theta = 0$ and $V_{ss} = V_{,\chi\chi} = m^2$.

While the curvature perturbations freeze, the entropy perturbations do not due to the presence of the potential term $V_{ss}$ in their equation of motion.  To be specific, if the potential term dominates over the Hubble friction term, the entropy perturbations (for the case $V_{,\chi\chi} = m^2$) will oscillate with constant amplitude.  On the other hand, if the Hubble friction term dominates, it is easy to check that the entropy perturbations freeze, just like the curvature perturbations.  Typically, we expect that the potential term will dominate first, and then the Hubble friction term will become dominant when $H > m$.  In this case, there will be a number of oscillations in the entropy perturbations before they freeze. Depending on where in the phase of the oscillation $\de s$ freezes, the entropy perturbations can be significantly damped, but generically the amplitude of $\de s$ before and after the $V_{ss}$ generated oscillations will be of the same order of magnitude.

Another possibility arises if the potential $V(\phi, \chi)$ changes its form for positive $\phi$.  For example, if $V(\phi, \chi) = \tf{4}{\pi} \arctan(\phi / \alpha) \cdot \tf{1}{2} m^2 \chi^2 - V_o e^{c \phi}$ (for some constant $\alpha$) then the potential in $\chi$ flips over at $\phi=0$, and the amplitude of the entropy perturbations will be exponentially amplified, rather than oscillating with a constant amplitude.  As a result, the evolution of the entropy perturbations during the ekpyrotic phase strongly depends on the choice of the potential $V(\phi, \chi)$.

In order to be conservative here we will assume that in this case $V_{,\chi\chi} = m^2$ and then the entropy perturbations oscillate with a constant amplitude, assuming the Hubble rate is sufficiently small so that the Hubble friction term in the equation of motion for $\de s$ remains negligible.  However, it is important to keep in mind that it is possible to choose potentials that can either amplify or damp (potentially by a very large factor) the entropy perturbations during the ekpyrotic phase of contraction.

Note that during ekpyrotic contraction the Hubble rate grows continuously, so ---assuming the potential \eqref{mb-pot} during the ekpyrotic contraction--- at some point the Hubble friction term will become the dominant term in \eqref{eom-ds} and at this time the long-wavelength entropy perturbations will freeze.  For this reason, as explained in detail in Sec.~\ref{section:bounce}, entropy perturbations will be amplified across the bounce in this cosmological scenario only if (i) the ekpyrotic phase is relatively short and ends before it would otherwise cause the entropy perturbations to freeze, or (ii) $V_{ss} \neq 0$ during the bounce phase and/or the surrounding kinetic phases.  (Note that for different potentials, a long ekpyrotic phase is possible if $V_{ss}$ grows faster than $H^2$ during the ekpyrotic phase.)

In condition (i), the short ekpyrotic phase could either be followed by a kinetic contracting phase if the ekpyrotic potential flattens out, or immediately by the bounce: in both cases, a short ekpyrotic contracting phase can freeze the curvature perturbations (recall that the long-wavelength curvature perturbations freeze at the onset of the ekpyrotic phase), but not the entropy perturbations. This is precisely the condition that must be met if $V_{ss} = 0$ during the bounce in order to obtain an amplification of the entropy perturbations across the bounce while leaving the curvature perturbations untouched.

For the case that $V_{ss}$ is not negligible during the bounce or surrounding kinetic phases, the relation $\dot{\de s} \neq 0$ generically holds for the simple reason that in this case (except for some very specific choices of the potential) neither of the two solutions to \eqref{eom-ds} is a constant function. Hence, even if the entropy perturbations are frozen before the bounce, their amplitude will be modified by the non-vanishing $V_{ss}$ term.  (On the other hand, the curvature perturbations remain frozen so long as $\dot \theta = 0$.)  In this case, the effect on the amplitude of $\de s$ strongly depends on the form of the potential.  In particular, if the transverse potential is unstable the entropy perturbations can be significantly amplified, no matter their initial conditions at the onset of the bounce phase.

In the remainder of this section we will consider the other possibility that allows for the entropy perturbations to grow across the bounce, namely the case where on the one hand the ekpyrotic phase is sufficiently long so that the curvature perturbations freeze, but on the other hand sufficiently short so that the entropy perturbations do not freeze.  Then, if this short ekpyrotic phase is followed by a kinetic-dominated phase (or perhaps immediately followed by the bounce phase), the non-trivial time-dependence of the entropy perturbations at the transition time between the ekpyrotic and kinetic-dominated phases will lead to a growth in the amplitude of the entropy perturbations across the bounce, while the the curvature perturbations remain frozen.

\begin{figure} 
	\centering
	\includegraphics[width=0.5\textwidth]{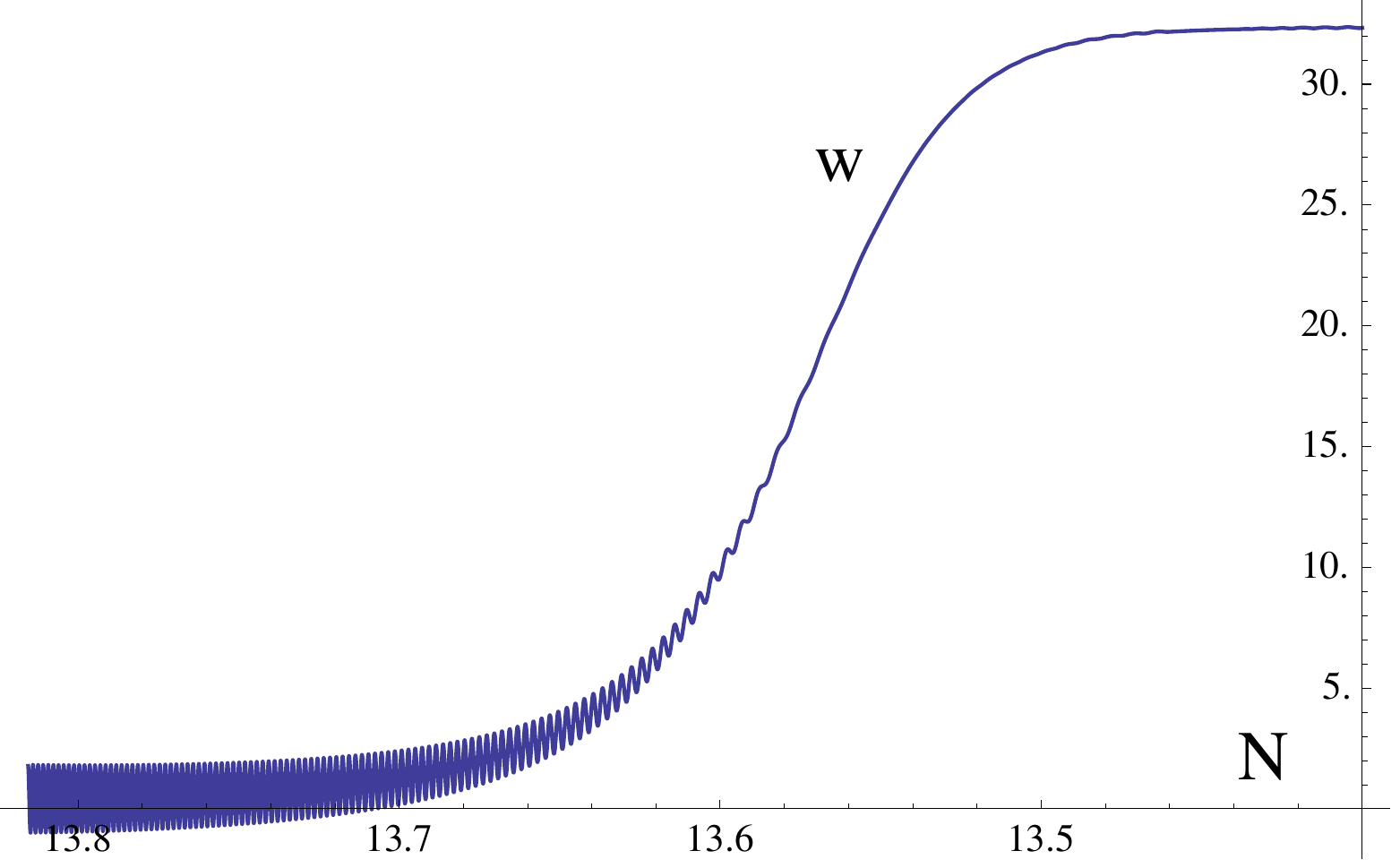}
	\caption{\footnotesize \hangindent=10pt
Evolution of the effective equation of state of the background, plotted with respect to the e-fold time $d\mN = d \ln a = H dt$ (which decreases in a contracting universe). The transition between the matter-dominated and ekpyrotic phases of contraction occurs when the effective equation of state stops oscillating around 0 and grows to a large constant value. In this case, while this is not a sharp transition, the transition begins at approximately $\mN = 13.7$.}
	\protect
	\label{fig:m-eos}
\end{figure}

\begin{figure}[htbp]
	\begin{minipage}{0.5\textwidth}
		\includegraphics[width=0.92\textwidth]{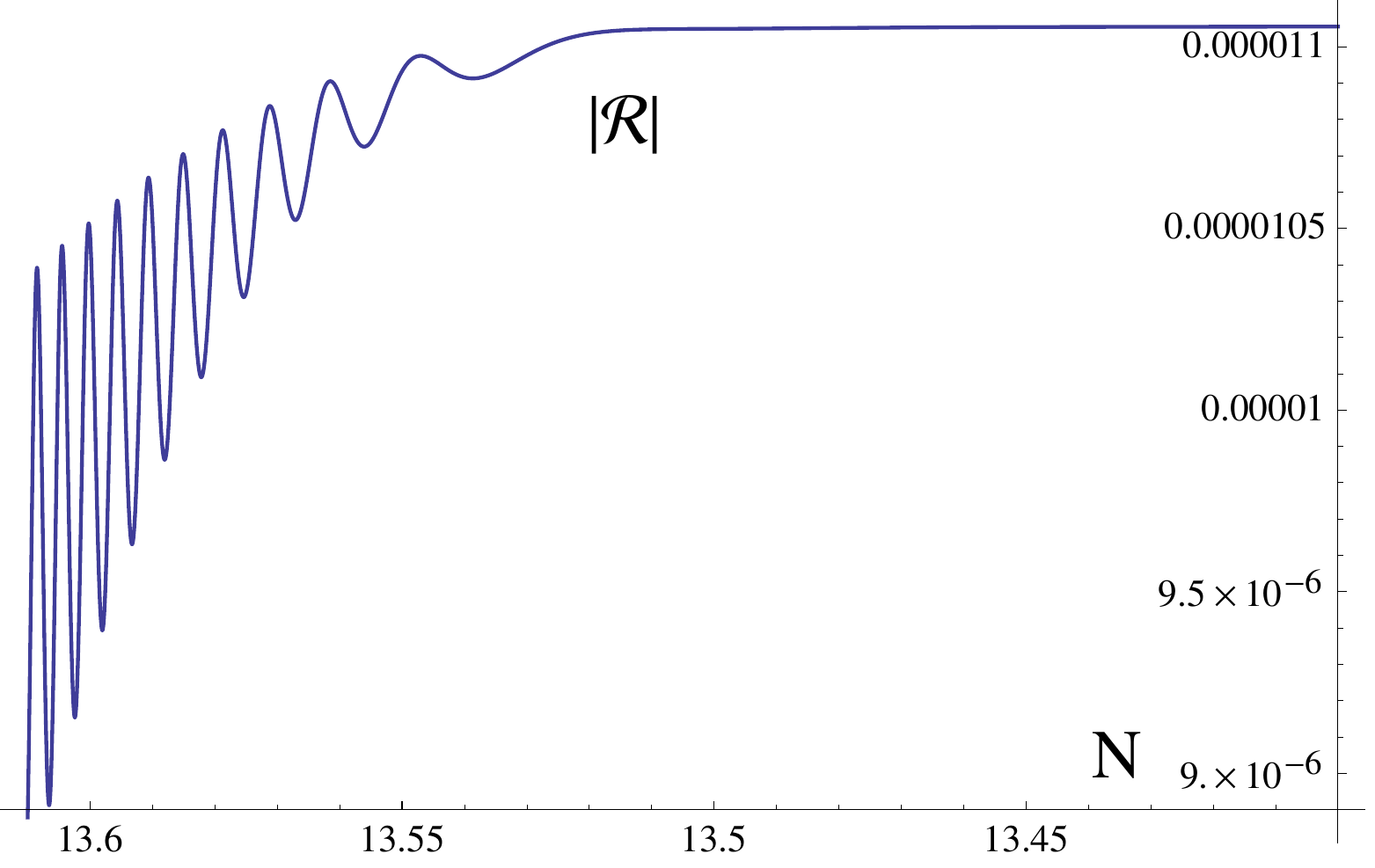}
	\end{minipage}%
	\begin{minipage}{0.5\textwidth}
		\includegraphics[width=0.92\textwidth]{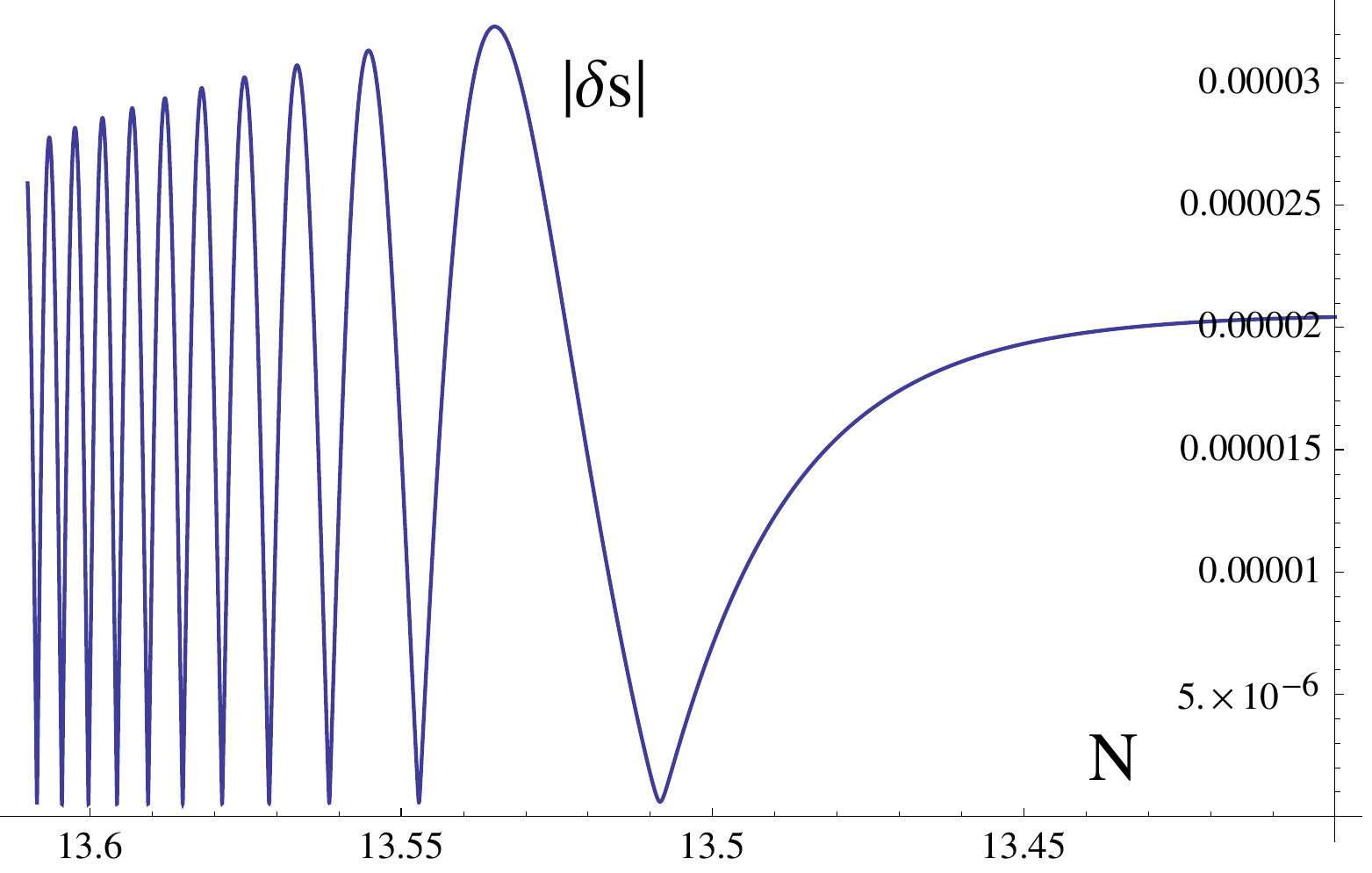}
	\end{minipage}%
	\caption{\footnotesize \hangindent=10pt
A numerical solution of the evolution of the curvature and entropy perturbations during the matter and ekpyrotic phases of contraction. The left panel shows the evolution of the curvature perturbations which oscillate with a growing amplitude during the matter phase and then freeze during the ekpyrotic phase when $\dot\theta$ goes to zero. The right panel shows the evolution of the entropy perturbations which also oscillate with a growing amplitude during the matter phase, and continue to have a non-trivial evolution during the ekpyrotic phase so long as the Hubble rate remains smaller than $m$.  Note that near the equality time $t_e$ ---i.e.~near the transition between the matter and ekpyrotic phases of contraction--- the amplitudes of the curvature and entropy perturbations are of the same order of magnitude, as expected from \eqref{same-amp}.}
	\label{fig:Rds}
\end{figure}

The dynamics described above can be solved numerically.  As a specific example, we assume the initial conditions
\bea
& \chi(t_o)=1.74 \times 10^{-4} \, ,\quad \dot \chi(t_o)=1.44 \times 10^{-8} \, , \quad
\phi(t_o)=-\frac{3}{2} \, ,\quad \dot \phi(t_o)= 10^{-9}\,, & \\
& a(t_o)=10^6 \, ,\quad m=10^{-6} \, , \quad V_0=-10^{-10} \, ,\quad c=10 \, , & \nn
\eea
during the matter-dominated phase of contraction for the background variables. Solving the differential equations numerically in terms of the e-folding time $d\mN = d \ln a = H dt$, it is possible to determine the effective equation of state, which is plotted in Fig.~\ref{fig:m-eos}.  During the matter-dominated phase of contraction, the effective equation of state initially oscillates rapidly around 0 (vanishing on average) and then its sharp increase to a large constant value signals the onset of the ekpyrotic phase.

The initial conditions for the numerical evolution of the perturbations are taken to be
\bea
& \mR_{k_o}=7.75 \times 10^{-8}  \, , \quad
\dot \mR_{k_o}= 9.41\times  10^{-14}  -i \cdot 7.75 \times  10^{-13} & \\
& \delta s_{k_o}=8.45 \times 10^{-6} \, , \quad
\dot \delta s_{k_o}=4.23\times  10^{-14} -i \cdot 5.92 \times  10^{-14} \, , & \nn
\eea
for the $k_o = 7 \times 10^{-3}$ Fourier mode.  (Note that while $\mR, \de s \in \mathbb{R}$, their Fourier modes $\mR_k, \de s_k \in \mathbb{C}$ and therefore these initial conditions are complex-valued.)  These initial conditions for the sub-horizon modes during the matter-dominated contracting phase are set following \eqref{born-zero}; the result of the numerics shown in Fig.~\ref{fig:Rds} demonstrates how the mixing between the curvature and entropy perturbations ensures that their amplitudes be of the same order of magnitude at the time of equality between the matter-dominated and ekpyrotic phases and is in good agreement with the result \eqref{born-exp}.  Indeed, the numerics show that both the curvature and entropy perturbations are initially oscillating with amplitudes that are increasing at different rates. Then, shortly after the transition to the ekpyrotic phase ---shown in Fig.~\ref{fig:m-eos} to occur around $\mN = 13.7$ for these initial conditions--- the curvature perturbations freeze when $\dot\theta = 0$ (in this case at $\mN \approx 13.5$, which coincides with the time when the effective equation of state becomes constant) while the entropy perturbations continue to have a non-trivial (i.e.~$\dot{\delta s} \neq 0$) evolution after the onset of the ekpyrotic contraction phase.

The key points here are the following: (i) at the beginning of the ekpyrotic phase, the curvature and entropy perturbations are of the same order of magnitude; (ii) the curvature perturbations freeze shortly after the onset of the ekpyrotic phase; (iii) while $\dot{\delta s} \neq 0$ so long as $|V_{ss}| > H^2$.   If the ekpyrotic phase is sufficiently short and ends before the Hubble rate becomes comparable to $m$ (at which point the entropy perturbations would freeze), then in the next phase the entropy perturbations will necessarily have a time-dependent term which will grow during the bounce and kinetic-dominated phases.

Finally, during the bounce ---as explained in Sec.~\ref{section:bounce}--- the curvature perturbations remain frozen, while the amplitude of the entropy perturbations can increase by two or three orders of magnitude depending on the transverse potential $V_{ss}$ (although only in the case that the entropy perturbations are not frozen in the pre-bounce phase).  Even if, as is assumed here, the transverse potential does not play a significant role, the amplitude of the entropy perturbations can increase by one or two orders of magnitude once the kinetic-dominated post-bounce phases are included.  (And if there is an unstable transverse potential, the increase can be significantly greater.)  So, after the bounce, the entropy perturbations are expected to have a significantly larger amplitude than the curvature perturbations, by two or three orders of magnitude.

\subsection{Conversion After the Bounce}
\label{ss.conv-mb}

Converting the entropy perturbations to curvature perturbations before the bounce does not significantly change the predictions of this cosmological scenario without any conversion of entropy perturbations for the simple reason that, since before the bounce entropy perturbations are of the same order of magnitude as curvature perturbations, they can at most approximately double the amplitude of the curvature perturbations.

However, if the entropy perturbations are converted to curvature perturbations after the bounce, the situation is quite different.  In this case, the entropy perturbations will in fact be the dominant contribution to the curvature perturbations (assuming an efficient conversion) and lead to a much greater amplitude of curvature perturbations than would be the case otherwise.  To be specific, in the case that the entropy perturbations are converted to curvature perturbations after the bounce, it is sufficient to have an amplitude of $\Delta_\mR(k)^2 = k^3 |\mR_k|^2 / 2 \pi^2 \sim 10^{-11}$ or even smaller before the bounce.  This is viable in this scenario since the entropy perturbations will be of the same amplitude before the bounce and are amplified by at least one order of magnitude during the bounce phase before being converted to curvature perturbations, giving an amplitude consistent with observations.  Importantly, as we shall now explain, this scenario with a conversion of entropy perturbations to curvature perturbations after the bounce predicts a small tensor-to-scalar ratio and small but non-negligible non-Gaussianities if the conversion process is efficient.

In the simplest matter bounce models, the tensor-to-scalar ratio is typically predicted to be too large \cite{Quintin:2015rta} (although a small tensor-to-scalar ratio is possible if there are multiple matter fields \cite{Cai:2011zx} or if the matter field has a small sound speed \cite{Cai:2014jla}; another possibility is that a loop quantum cosmology bounce suppresses the amplitude of tensor modes with respect to scalar modes \cite{WilsonEwing:2012pu}).  The reason for this is that the equation of motion for the tensor modes $h = \mu / a$ in Fourier space,
\be
\mu'' + \left( k^2 - \fr{a''}{a} \right) \mu = 0,
\ee
generates essentially the same dynamics in the tensor mode $h$ as the dynamics for $\mR$.  In particular, assuming that the tensor modes originate in their vacuum quantum state, at the end of the matter-dominated phase of contraction, their amplitude is also determined by the Hubble rate at the equality time $H_e$ \cite{Cai:2013kja},
\be
h_k(t_e) \sim H_e \, k^{-3/2}.
\ee
Then, the tensor perturbations freeze during the ekpyrotic phase of contraction, just like the curvature perturbations.  It follows that the tensor-to-scalar ratio
\be
r = \fr{\Delta_h(k)^2}{\Delta_\mR(k)^2}
\ee
is of the order of unity before the bounce in this scenario, while observations of the CMB give the bound $r < 0.12$ (95\% CL) \cite{Ade:2015tva}.

However, if the amplitude of the entropy perturbation $\delta s$ grows by at least an order of magnitude during the bounce and becomes the dominant contribution to the curvature perturbation $\mR$ following an efficient conversion process after the bounce, this increases the amplitude of curvature perturbations with respect to tensor perturbations and suppresses the tensor-to-scalar ratio by at least 2 orders of magnitude, giving a value of $r$ well below the observational constraint.  In this cosmological model with a conversion of the entropy perturbations after the bounce, the exact numerical prediction of the tensor-to-scalar ratio is strongly model-dependent.  However, if the entropy perturbations $\delta s$ are significantly amplified (say by two orders of magnitude or more), then we expect $r < 10^{-4}$.

Furthermore, non-Gaussianities are also \emph{a priori} expected to be small if the entropy perturbations are converted to curvature perturbations after the bounce in this scenario. With our assumed form of the potential in Eq. \eqref{mb-pot}, on average the third derivative in the transverse/entropic direction is small, and thus we may expect the intrinsic non-Gaussianity in the entropy perturbations to be small. Then, if the entropy perturbations are the dominant contribution to curvature perturbations at first order in perturbation theory, the dominant contribution to the higher order terms in the curvature perturbation will also likely come from converted entropy perturbations, and then the dominant contribution to the non-Gaussianities in fact typically arises from the conversion process itself, not from the non-Gaussianities in the entropy perturbations. This is just as we have discussed in Sec. \ref{sec:ek_afterbounce} for ekpyrotic models, and we then expect to find the same result, namely that for efficient conversions the final non-Gaussianities in the curvature perturbation are small ($|f_{NL}|\lesssim 5$) and in agreement with observational bounds \cite{Ade:2015ava}.  While these arguments are indicative that non-Gaussianities are predicted to be small here, we leave for future work the more detailed analysis that is needed in order to confirm these expectations.

Note that in the case that the entropy and curvature perturbations are of the same order, then both will contribute to the non-Gaussianities.  The standard matter bounce scenario predicts a distinctive form of non-Gaussianities \cite{Cai:2009fn}, and it would be interesting to see how these combine with the non-Gaussianities coming from the conversion process in the case that entropy perturbations are of a similar amplitude.  This question is also left for future work.

%%%%%%%%%%%%%%%%%%%%%%%%%%%%%%%%%%%%%%%%%%%%%%%%%%%%%%%%%%
\section{Discussion}
\label{section:discussion}

We have analysed ekpyrotic and matter-dominated contracting cosmologies in which entropy perturbations provide the dominant source of the final curvature perturbations, and thus the dominant source of the eventual temperature fluctuations observed in the cosmic microwave background. In these models entropy perturbations are generated during the contracting phase. However, in contrast to earlier models, these entropy perturbations are not converted into curvature perturbations at the end of the contracting phase, but only during the subsequent expanding phase, following a bounce. The crucial difference in having the entropy perturbations around for longer is that they become increasingly important. We have shown that this is due to a combination of effects: (i) they grow by a factor of a few during the bounce itself, i.e.~during the period when the null energy condition is violated, (ii) they grow logarithmically during the kinetic phase that immediately follows the bounce (and will also grow logarithmically during any kinetic phase that may precede the bounce), and (iii) entropy perturbations can potentially grow significantly more if there is an unstable transverse potential at any time during the ekpyrotic, kinetic or bounce phases.  An important point is that although (in the absence of an unstable potential) the growth is only logarithmic during the kinetic-dominated phases, the cumulated growth can be very significant and can in fact increase the amplitude of the entropy perturbations by up to several orders of magnitude.

Thus if a conversion event of entropy into curvature fluctuations occurs after the bounce, then typically one may expect the entropy perturbation to be the main originator of the late-time density fluctuations. Note that there is no conflict with the fact that the temperature fluctuations in the CMB are observed to be adiabatic, as in the present models the entropy perturbations are converted into curvature perturbations during the expanding phase, while we assume that the universe subsequently reaches thermal equilibrium at which time the remaining entropy perturbations vanish. 

The extra amplification of entropy perturbations implies that the fluctuations generated during the contracting phase can have a smaller initial amplitude than what is currently assumed in such models, i.e.~the contracting phase can end at a lower energy scale. The main observational signature of the conversion process are the associated non-Gaussianities. While they can be small as long as the conversion process is efficient, they are certainly not negligible. Our numerical studies suggest that efficient conversions typically lead to 
\be
-5 \lesssim f_{NL} \lesssim +5\,,
\ee
which is consistent with the current observational bounds $f_{NL} = 0.8 \pm 10.0$ at the $95\%$ confidence level from the Planck collaboration \cite{Ade:2015xua}, but within the reach of not-too-distant future experiments such as the SKA \cite{Gaensler}. A further distinguishing feature of these models is that one expects them to also have a non-trivial running of the spectral index, as shown in \cite{Lehners:2015mra}. Therefore, these models provide interesting alternatives to existing models, with the benefit that they can be tested by near-future observations.

As for most cosmological models, a challenge remains to embed the present models in a more fundamental framework.  There are several aspects to this challenge: an obvious one is to find a model that contains the appropriate fields and potentials.  While it should be straightforward to find potentials that lead to a conversion process \cite{Lehners:2007nb}, it remains to be demonstrated that one can derive potentials from a more fundamental theory that lead to efficient conversions. As we have discussed, the efficiency of the conversion process is paramount in obtaining reliable and phenomenologically interesting predictions.  What is more, non-singular bounces are extremely interesting effective theories, but it remains an open problem to derive them as effective descriptions of say loop quantum gravity bounces, or from a string theory setting. (Bounces naturally arise in loop quantum cosmology \cite{Ashtekar:2006wn} as well as in condensate states of loop quantum gravity \cite{Oriti:2016ueo}, but it is not known if loop quantum gravity effects can cause an efficient conversion process near the bounce point; bounces can also be embedded in supergravity \cite{Koehn:2013upa}, but a possible relation to a more fundamental theory remains unknown --- these are both promising starting points, but much remains to be done.) Also, as we mentioned in discussing a braneworld inspired setting, the conversion process may have interesting links to the formation and abundance of dark matter. Given the large number of cosmological models that can potentially explain the origin of the primordial fluctuations, it might well be the case that our trust or distrust of various models will come not only from more detailed observations (which will certainly play a leading role), but also from additional correlations with altogether different observations, such as the dark matter abundance. This particular point thus deserves further study.

\acknowledgments

We express our gratitude to the Max Planck Society for its support of the Theoretical Cosmology group at the Albert Einstein Institute. This work was supported in part by a grant from the John Templeton Foundation.
JLL would like to thank Perimeter Institute for hospitality while this work was completed. This research was also supported in part by the Perimeter Institute for Theoretical Physics. Research at Perimeter Institute is supported by the Government of Canada through Industry Canada and by the Province of Ontario through the Ministry of Economic Development \& Innovation.

\raggedright

\end{document}